\documentclass[aps,showpacs,preprintnumbers]{revtex4}

\usepackage{amsfonts}

\usepackage{graphicx}% Include figure files
\usepackage{dcolumn}% Align table columns on decimal point
\usepackage{bm}% bold math
\usepackage{epsfig}
\newcommand{\vsig}{\mbox{\boldmath$\sigma$\unboldmath}}

\begin{document}

\title{Strong decays of heavy-light mesons in a chiral quark model}
\author{
Xian-Hui Zhong$^{1,3}$ \footnote {E-mail: zhongxh@ihep.ac.cn} and
Qiang Zhao$^{1,2,3}$ \footnote {E-mail: zhaoq@ihep.ac.cn}}

\affiliation{1) Institute of High Energy Physics,
       Chinese Academy of Sciences, Beijing 100049, P.R. China
}
\affiliation{2) Department of Physics, University of Surrey,
Guildford, GU2 7XH, United Kingdom
            }

\affiliation{3) Theoretical Physics Center for Science Facilities,
CAS, Beijing 100049, P.R. China}

%\date{\today}

\begin{abstract}

We carry out a systematic study of the heavy-light meson  strong
decays in a chiral quark model. For the $S$-wave vectors
($D^*(2007)$, $D^{*\pm}(2010)$), $P$-wave scalars ($D^*_0(2400)$,
$B^*_0(5730)$) and tensors ($D^*_2(2460)$, $D^*_{s2}(2573)$), we
obtain results in good agreement with the experimental data. For
the axial vectors $D_1(2420)$ and $D_1^\prime(2430)$, a state
mixing scheme between $1^1P_1$ and $1^3P_1$ is favored with a
mixing angle $\phi\simeq-(55\pm 5)^\circ$, which is consistent
with previous theoretical predictions. The same mixing scheme also
applies to $D_{s1}(2460)$ and $D_{s1}(2536)$ that accounts for the
narrow width of the $D_{s1}(2536)$ and its dominant decay into
$D^*K$. For $B_1(5725)$ and $B_1^\prime(5732)$, such a mixing
explains well the decay width of the former but leads to an even
broader $B_1^\prime(5732)$. Predictions for the strange-bottom
axial vectors are also made. For the undetermined meson
$D^*(2640)$, we find that they fit in well the radially excited
state $2^3S_1$ according to its decay mode. The newly observed
$D^*_{sJ}(2860)$ strongly favors the $D$-wave excited state
$1^3D_3$. For $D^*_{sJ}(2632)$ and $D^*_{sJ}(2690)$, we find they
are difficult to fit in any $D_s$ excitations in that mass region,
if the experimental data are accurate. Theoretical predictions for
decay modes of those unobserved states as multiplets of $2S$ and
$1D$ waves are also presented, which should be useful for further
experimental search for those states.

\end{abstract}


\maketitle

\section{Introduction}

During the past several years, significant progresses have been
made in the observation of the heavy-light mesons. In 2003, two
new narrow charm-strange mesons $D^*_{sJ}(2317)$ and
$D^*_{sJ}(2460)$ were observed by BaBar, CLEO and Belle
\cite{Aubert:2003fg,Aubert:2003pe,Besson:2003cp,Abe:2003jk,Krokovny:2003zq}.
Recently, BaBar reported another two new charm-strange mesons,
i.e. $D^*_{sJ}(2860)$ with a width of $(47\pm 17)$ MeV and
$D^*_{sJ}(2690)$ with a width of $(112\pm 43)$ MeV in the $DK$
decay channel~\cite{Aubert:2006mh}. Meanwhile, Belle reported a
new vector state $D^*_{sJ}(2708)$ with a width of $(108\pm
23^{+36}_{-31})$ MeV \cite{:2007aa}. The $D^*_{sJ}(2690)$ and
$D^*_{sJ}(2708)$ are believed to be the same state since their
masses and widths are consistent with each other. In the $B$ meson
sector two narrow states $B_1(5725)$ and $B^*_2(5740)$ were
reported by CDF \cite{B:2007CDF}, and are assigned as orbitally
excited $B$ mesons. They were confirmed by D0 collaboration with
slightly different masses \cite{Abazov:2007vq}. CDF collaboration
also reported their strange analogues, $B_{s1}(5829)$ and
$B^*_{s2}(5840)$, as orbitally excited $B_s$
mesons~\cite{:2007tr}. The $B^*_{s2}(5840)$ is also observed by D0
collaboration \cite{:2007sna}. About the recent experimental
status of the heavy-light mesons, many reviews can be found in
Refs.~\cite{Bianco:2005hj,Kravchenko:2006qx,Waldi:2007bp,Zghiche:2007im,
Poireau:2007cb,Mommsen:2006ai,Kreps:2007mz}.

To understand the nature of the heavy-light mesons, especially the
newly observed states, and to establish the heavy-light meson
spectroscopy, a lot of efforts have been made on both experiment
and theory. For example, one can find recent discussions about the
dynamics and decay properties of the heavy-light mesons given by
Close and Swanson \cite{Close:2005se}, Godfrey
\cite{Godfrey:2005ww}, and other previous analyses in
Refs.~\cite{Godfrey:1985xj,Godfrey:1986wj,Isgur:1991wq,Di
Pierro:2001uu,Falk:1995th,Eichten:1993ub,Henriksson:2000gk,Tregoures:1999du,Goity:1998jr,
Zhu:1998wy,Dai:1998ve,Orsland:1998de}. For the new observed
heavy-light mesons, such as $D^*_{sJ}(2860)$ and $D^*_{sJ}(2690)$,
various attempts on the explanation of their nature have been made
\cite{vanBeveren:2006st,Zhang:2006yj,Wei:2006wa,Close:2006gr,Colangelo:2007ds,
Wang:2007nfa,Guo:2006rp,Faessler:2008vc,Yasui:2007dv,Colangelo:2006rq,Koponen:2007nr}.
Many systematic studies are devoted to establish the $D$, $D_s$,
$B$, and $B_s$
spectroscopies~\cite{Vijande:2004he,Vijande:2006hj,Matsuki:2007zz,Vijande:2007ke,Li:2007px,Swanson:2007rf},
while some earlier works can be found in
Refs.~\cite{Godfrey:1985xj,Ebert:1997nk}. Recent reviews of the
status of the theory study of the heavy-light mesons can be found
in
Refs.~\cite{Barnes:2005zy,Swanson:2005tq,Swanson:2006st,Rosner:2006jz,Zhu:2007xb,Rosner:2006sv,
Nicotri:2007in,Klempt:2007cp}

On the one hand, the improved experimental measurements help
clarify some old questions on the spectrum. On the other hand,
they also raise some new ones which need further experimental and
theoretical studies \cite{Colangelo:2007ur,Ananthanarayan:2007sa}.
For instance, $D^*(2640)$ reported by DELPHI in
$D^{*+}\pi^+\pi^-$~\cite{Abreu:1998vk} as the first radial excited
state still has not yet been confirmed by any other experiments.
The spin-parity of the narrow $D_1(2420)$ also need confirmations.
The status of the broad $D^*_0(2400)$ is not clear at all, its
measured mass and width have too large uncertainties. For the
$D_s$ spectroscopy, the low masses of the $D^*_{sJ}(2317)$ and
$D^*_{sJ}(2460)$ still cannot be well explained by theory; whether
they are exotic states is an open question. Theoretical
predictions for the $D^*_{sJ}(2860)$ and $D^*_{sJ}(2690)$ are far
from convergence. The narrow state $D^*_{sJ}(2632)$ seen by SELEX
Collaboration~\cite{Evdokimov:2004iy} cannot be naturally
explained by any existed theory. Nevertheless, since the flavor
symmetry of the heavy-light mesons is badly broken, mixture of
states with the same $J^P$ may occur. This will add further
complexities into the meson spectrum and further theoretical
investigations are needed.

In this work, we make a systematic study of the strong decays of
heavy-light mesons in a chiral quark model. In the heavy-quark
infinite mass limit, the flavor symmetry does no longer exist in
the heavy-light mesons, which allows us to describe the initial
and final $D$, $D_s$, $B$, and $B_s$ mesons in a nonrelativistic
framework self-consistently. The meson decay will proceed through
a single-quark transition by the emission of a pseudoscalar meson.
An effective chiral Lagrangian is then introduced to account for
the quark-meson coupling. Since the quark-meson coupling is
invariant under the chiral transformation, some of the low-energy
properties of QCD are retained. This approach is similar to that
used in Refs.~\cite{Godfrey:1985xj,Godfrey:1986wj}, except that
the two constants in the decay amplitudes of
Refs.~\cite{Godfrey:1985xj,Godfrey:1986wj} are replaced by two
energy-dependent factors deduced from the chiral Lagrangian in our
model.

The chiral quark model approach has been well developed and widely
applied to meson photoproduction
reactions~\cite{Manohar:1983md,qk1,qk2,qkk,Li:1997gda,qkk2,qk3,qk4,qk5}.
Its recent extension to describe the process of $\pi N$ scattering
and investigate the strong decays of charmed baryons also turns
out to be successful and
inspiring~\cite{Zhong:2007fx,Zhong:2007gp}.

The paper is organized as follows. In the subsequent section, the
heavy-light meson in the quark model is outlined. Then, the
non-relativistic quark-meson couplings are given in Sec.\
\ref{qmc}. The decay amplitudes are deduced in Sec.\ \ref{apt}. We
present our calculations and discussions in Sec.\ \ref{cd}.
Finally, a summary is given in Sec.\ \ref{sum}.

\section{meson spectroscopy }

\subsection{Harmonic oscillator states}

For a heavy-light $\bar{Q}q$ system consisting light quark 1 and
heavy quark 2 with masses $m_1$ and $m_2$, respectively, its
eigen-states are conventionally generated by a harmonic oscillator
potential
\begin{eqnarray} \label{hm1}
\mathcal{H}=\frac{1}{2m_1}\mathbf{p}^2_1+\frac{1}{2m_2}\mathbf{p}^2_2+
\frac{3}{2}K(\mathbf{r}_1-\mathbf{r}_2)^2,
\end{eqnarray}
where vectors $\textbf{r}_{j}$ and $\textbf{p}_j$ are the
coordinate and momentum for the $j$-th quark in the meson rest
frame, and $K$ describes the oscillator potential strength which
is independent of the flavor quantum number. One defines the
Jacobi coordinates to eliminate the c.m. variables:
\begin{eqnarray}
\mathbf{r}&=&\mathbf{r}_1-\mathbf{r}_2,\label{zb1}\\
\mathbf{R}_{c.m.}&=&\frac{m_1\mathbf{r}_1+m_2\mathbf{r}_2}{m_1+m_2}\label{zb3}.
\end{eqnarray}
With the above relations (\ref{zb1}--\ref{zb3}), the oscillator
hamiltonian (\ref{hm1}) is reduced to
\begin{eqnarray} \label{hm2}
\mathcal{H}=\frac{P^2_{cm}}{2 M}+\frac{1}{2\mu}\mathbf{p}^2+
\frac{3}{2}Kr^2.
\end{eqnarray}
where
\begin{eqnarray} \label{mom}
\mathbf{p}=\mu\dot{\mathbf{r}},\ \ \mathbf{P}_{c.m.}=M
\mathbf{\dot{R}}_{c.m.},
\end{eqnarray}
with
\begin{eqnarray} \label{mass}
M=m_1+m_2,\ \ \mu=\frac{m_1 m_2}{m_1+m_2}.
\end{eqnarray}
From Eqs.(\ref{zb1}--\ref{zb3}) and (\ref{mom}), the coordinate
$\mathbf{r}_j$ can be expressed as functions of the Jacobi
coordinate $\textbf{r}$:
\begin{eqnarray}
\mathbf{r}_1&=&\mathbf{R}_{c.m.}+\frac{\mu}{m_1}\mathbf{r},\\
\mathbf{r}_2&=&\mathbf{R}_{c.m.}-\frac{\mu}{m_2}\mathbf{r},
\end{eqnarray}
and the momentum $\mathbf{p}_j$ is given by
\begin{eqnarray}
\mathbf{p}_1&=&\frac{m_1}{M}\mathbf{P}_{c.m.}+\mathbf{p},\\
\mathbf{p}_2&=&\frac{m_2}{M}\mathbf{P}_{c.m.}-\mathbf{p}.
\end{eqnarray}

Using standard notation, the principal quantum numbers of the
oscillator is $N=(2n+l)$, the energy of a state is given by
\begin{eqnarray}
E_N&=&(N+\frac{3}{2})\omega,
\end{eqnarray}
and the frequency of the oscillator is
\begin{eqnarray}\label{freq}
\omega=(3K/\mu)^{1/2}.
\end{eqnarray}
In the quark model the useful oscillator parameter is defined by
\begin{eqnarray} \label{par}
\alpha^2=\mu \omega=\sqrt{\frac{2m_2}{m_1+m_2}}\beta^2,
\end{eqnarray}
where $\beta$ is the often used harmonic oscillator parameter with
a universal value $\beta=0.4$ GeV. Then, the wave function of an
oscillator is give by
\begin{eqnarray}
\psi^{n}_{l m}=R_{n l}Y_{l m}.
\end{eqnarray}

\begin{table}[ht]
\caption{The total wave function for the heavy-light mesons,
denoted by $|n ^{2S+1} L_{J}\rangle$. The Clebsch-Gordan series
for the spin and angular-momentum addition $|n ^{2S+1} L_{J}
\rangle= \sum_{m+S_z=J_z} \langle Lm,SS_z|JJ_z \rangle \psi^n_{Lm}
\chi_{S_z}\Phi $ has been omitted, where $\Phi$ is the flavor wave
function.} \label{wfS}
\begin{tabular}{|c|c|c|c|c|c|c }\hline\hline
$|n ^{2S+1} L_J\rangle$ & $J^P$ &\ \ wave function   \\
\hline
1 $^1 S_0$ & $0^-$ & $\psi^{0}_{00}\chi^0\Phi$       \\
\hline
1 $^3 S_1$& $1^-$ & $\psi^{0}_{00}\chi^1_{S_z}\Phi$     \\
\hline
1 $^1 P_1$& $1^+$ &$\psi^{0}_{1m}\chi^0\Phi$      \\
\hline
1 $^3 P_0$& $0^+$ &$\psi^{0}_{1 m}\chi^1_{S_z}\Phi$      \\
\hline
1 $^3 P_1$& $1^+$ &$\psi^{0}_{1 m}\chi^1_{S_z}\Phi$      \\
\hline
1 $^3 P_2$& $2^+$ &$\psi^{0}_{1 m}\chi^1_{S_z}\Phi$      \\
\hline
2 $^1 S_0$& $0^-$ &$\psi^{1}_{00}\chi^0\Phi$      \\
\hline
2 $^3 S_1$& $1^-$ &$\psi^{1}_{00}\chi^1_{S_z}\Phi$      \\
\hline
1 $^1 D_2$& $2^-$ &$\psi^{0}_{2 m}\chi^0\Phi$      \\
\hline
1 $^3 D_1$& $1^-$ &$\psi^{0}_{2 m}\chi^1_{S_z}\Phi$      \\
\hline
1 $^3 D_2$& $2^-$ &$\psi^{0}_{2 m}\chi^1_{S_z}\Phi$      \\
\hline
1 $^3 D_3$& $3^-$ &$\psi^{0}_{2 m}\chi^1_{S_z}\Phi$      \\
\hline
\end{tabular}
\end{table}

\subsection{Spin wave functions}

The usual spin wave functions are adopted. For the spin-0 state,
it is
\begin{eqnarray}
\chi^0&=&\frac{1}{\sqrt{2}}(\uparrow\downarrow-\downarrow\uparrow),
\end{eqnarray}
and for the spin-1 states, the wave functions are
\begin{eqnarray}
\chi^1_{1}&=&\uparrow\uparrow,\ \ \ \chi^1_{-1}=\downarrow \downarrow ,\nonumber\\
\chi^1_{0}&=&\frac{1}{\sqrt{2}}
(\uparrow\downarrow+\downarrow\uparrow) .
\end{eqnarray}

We take the heavy-quark infinite mass limit as an approximation to
construct the total wave function without flavor symmetry. All the
wave functions up to $1D$ states are listed in Tab. \ref{wfS}.

\section{The quark-meson couplings }\label{qmc}

In the chiral quark model, the low energy quark-meson interactions
are described by the effective Lagrangian \cite{Li:1997gda,qk3}
\begin{eqnarray} \label{lg}
\mathcal{L}=\bar{\psi}[\gamma_{\mu}(i\partial^{\mu}+V^{\mu}+\gamma_5A^{\mu})-m]\psi
+\cdot\cdot\cdot,
\end{eqnarray}
where $V^{\mu}$ and $A^{\mu}$ correspond to vector and axial
currents, respectively. They are given by
\begin{eqnarray}
V^{\mu} &=&
 \frac{1}{2}(\xi\partial^{\mu}\xi^{\dag}+\xi^{\dag}\partial^{\mu}\xi),
\nonumber\\
 A^{\mu}
&=&
 \frac{1}{2i}(\xi\partial^{\mu}\xi^{\dag}-\xi^{\dag}\partial^{\mu}\xi),
\end{eqnarray}
with $ \xi=\exp{(i \phi_m/f_m)}$, where $f_m$ is the meson decay
constant. For the $SU(3)$ case, the pseudoscalar-meson octet
$\phi_m$ can be expressed as
\begin{eqnarray}
\phi_m=\pmatrix{
 \frac{1}{\sqrt{2}}\pi^0+\frac{1}{\sqrt{6}}\eta & \pi^+ & K^+ \cr
 \pi^- & -\frac{1}{\sqrt{2}}\pi^0+\frac{1}{\sqrt{6}}\eta & K^0 \cr
 K^- & \bar{K}^0 & -\sqrt{\frac{2}{3}}\eta},
\end{eqnarray}
and the quark field $\psi$ is given by
\begin{eqnarray}\label{qf}
\psi=\pmatrix{\psi(u)\cr \psi(d) \cr \psi(s) }.
\end{eqnarray}

From the leading order of the Lagrangian [see Eq.(\ref{lg})], we
obtain the standard quark-meson pseudovector coupling at tree level
\begin{eqnarray}\label{coup}
H_m=\sum_j \frac{1}{f_m}I_j
\bar{\psi}_j\gamma^{j}_{\mu}\gamma^{j}_{5}\psi_j\partial^{\mu}\phi_m.
\end{eqnarray}
where $\psi_j$ represents the $j$-th quark field in a hadron, and
$I_j$ is the isospin operator to be given later.

In the quark model, the non-relativistic form of Eq. (\ref{coup}) is
written as \cite{Zhong:2007fx,Li:1997gda,qk3}
\begin{eqnarray}\label{ccpk}
H^{nr}_{m}&=&\sum_j\Big\{\frac{\omega_m}{E_f+M_f}\vsig_j\cdot
\textbf{P}_f+ \frac{\omega_m}{E_i+M_i}\vsig_j \cdot
\textbf{P}_i \nonumber\\
&&-\vsig_j \cdot \textbf{q} +\frac{\omega_m}{2\mu_q}\vsig_j\cdot
\textbf{p}'_j\Big\}I_j \varphi_m,
\end{eqnarray}
where $\vsig_j$ corresponds to the Pauli spin vector of the $j$-th
quark in a hadron. $\mu_q$ is a reduced mass given by
$1/\mu_q=1/m_j+1/m'_j$, where $m_j$ and $m'_j$ stand for the masses
of the $j$-th quark in the initial and final hadrons, respectively.
For emitting a meson, we have $\varphi_m=\exp({-i\textbf{q}\cdot
\textbf{r}_j})$, and for absorbing a meson we have
$\varphi_m=\exp({i\textbf{q}\cdot \textbf{r}_j})$. In the above
non-relativistic expansions,
$\textbf{p}'_j=\textbf{p}_j-\frac{m_j}{M}\mathbf{P}_{c.m.}$ is the
internal momentum for the $j$-th quark in the initial meson rest
frame. $\omega_m$ and $\textbf{q}$ are the energy and three-vector
momentum of the light meson, respectively. The isospin operator
$I_j$ in Eq. (\ref{coup}) is expressed as
\begin{eqnarray}
I_j=\cases{ a^{\dagger}_j(u)a_j(s) & for $K^+$, \cr
a^{\dagger}_j(s)a_j(u) & for $K^-$,\cr a^{\dagger}_j(d)a_j(s) &
for $K^0$, \cr a^{\dagger}_j(s)a_j(d) & for $\bar{K^0}$,\cr
a^{\dagger}_j(u)a_j(d) & for $\pi^-$,\cr a^{\dagger}_j(d)a_j(u)  &
for $\pi^+$,\cr
\frac{1}{\sqrt{2}}[a^{\dagger}_j(u)a_j(u)-a^{\dagger}_j(d)a_j(d)]
& for $\pi^0$, \cr   \cos\theta
\frac{1}{\sqrt{2}}[a^{\dagger}_j(u)a_j(u)+a^{\dagger}_j(d)a_j(d)]\cr
-\sin\theta a^{\dagger}_j(s)a_j(s)& for $\eta$,}
\end{eqnarray}
where $a^{\dagger}_j(u,d,s)$ and $a_j(u,d,s)$ are the creation and
annihilation operators for the $u$, $d$ and $s$ quarks. Generally,
$\theta$ ranges from $\simeq 32^\circ \sim 43^\circ$ depending on
quadratic or line mass relation applied \cite{PDG}. In our
convention, $\theta=45^\circ$ corresponds to the mixing scheme of
Ref.~\cite{Close:2005se}. We applied the same value in order to
compared with Ref.~\cite{Close:2005se}. However, we note in advance
that within the commonly accepted range of $\theta$, our results do
not show great sensitivities due to the relatively large
uncertainties of the present experimental data.

%For the physical state of $\eta$ meson is the mixture of the SU(3)
%wave function $\eta_8$ and $\eta_1$, a typical quark model $\eta$
%mixing angle $\theta=45^\circ$ in the flavor basis is adopted in the
%calculations \cite{Close:2005se,PDG}.

\begin{table}[ht]
\caption{The spin-factors used in this work.} \label{gfactor}
\begin{tabular}{|l |l| l|c|c|c|c }\hline
\hline
$g^z_{10}=\langle \chi^0|\sigma_{1z}|\chi^1_{0}\rangle=1$  \\

$g^+_{10}=\langle \chi^0|\sigma^+_{1}|\chi^1_{-1}\rangle=\sqrt{\frac{1}{2}}$ \\

$g^-_{10}=\langle \chi^0|\sigma^-_{1}|\chi^1_{1}\rangle=-\sqrt{\frac{1}{2}}$ \\

$g^z_{01}=\langle \chi^1_{0}|\sigma_{1z}|\chi^0\rangle=1$   \\

$g^+_{01}=\langle \chi^1_{1}|\sigma^+_{1}|\chi^0\rangle=-\sqrt{\frac{1}{2}}$ \\

$g^z_{11}=\langle \chi^1_{1}|\sigma_{1z}|\chi^1_{1}\rangle=1$  \\

$g^+_{11}=\langle \chi^1_{1}|\sigma^+_{1}|\chi^1_{0}\rangle=\sqrt{\frac{1}{2}}$  \\
\hline

\end{tabular}
\end{table}

\begin{center}
\begin{table}[ht]
\caption{ The decay amplitudes for $|n \ ^{2S+1} L_{J}
\rangle\rightarrow |1^1S_0\rangle \mathbb{P}$. $g_I$ is a isospin
factor which is defined by $g_I=\langle
\phi_\Sigma|I_1|\phi_\Lambda\rangle$. In the Tab.
\ref{asa}--\ref{ase}, the overall factor $F(q')=\exp
\left(-\frac{q^{\prime 2}}{4\alpha^2}\right)$, which plays the role
of the decay form factor, is omitted for simplify, where $q^\prime=
(\mu/m_1)q$. In the tables, we have defined
$\mathcal{R}\equiv(\mathcal{G}q-\frac{1}{2}hq^\prime)$. Various
spin-factors used in this work are listed in the Tab.
\ref{gfactor}.} \label{asa}
\begin{tabular}{|c|c|c|c|c|c|c }\hline\hline
initial state                                                                             & amplitude\\
\hline
$1^3S_1  (1^-)    $                      & $g_Ig^z_{10}\mathcal{R} $\\
\hline
$1^1P_1  (1^+)    $              &forbidden\\
\hline
$1^3P_0  (0^+)    $              &$i\frac{1}{\sqrt{6}}g_Ig^z_{10}\mathcal{R}\frac{q^\prime}{\alpha} +i\frac{1}{\sqrt{6}}g_I(\sqrt{2}g^+_{10}+g^z_{10})h\alpha  $ \\
\hline
$1^3P_1   (1^+)   $                 &forbidden\\
\hline
$1^3P_2    (2^+)  $                 &$i\frac{1}{\sqrt{3}}g_Ig^z_{10}\mathcal{R}\frac{q^\prime}{\alpha} $\\
\hline
$2^1S_0    (0^-)    $                 &forbidden \\
\hline
$2^3S_1     (1^-)   $                 &$\frac{1}{\sqrt{24}}g_Ig^z_{10}\mathcal{R}(\frac{q^\prime}{\alpha})^2 +\sqrt{\frac{1}{6}}g_Ig^z_{10}hq^\prime  $ \\
\hline
$1^1D_2      (2^-)  $                  &forbidden\\
\hline
$1^3D_1      (1^-)                  $                   &$\frac{1}{\sqrt{30}}g_Ig^z_{10}\mathcal{R}(\frac{q^\prime}{\alpha})^2 +\sqrt{\frac{3}{5}}g_Ig^+_{10}hq^\prime  $\\
\hline
$1^3D_2      (2^-)                 $                   &forbidden\\
\hline
$1^3D_3      (3^-)                  $                   & $-\frac{1}{\sqrt{20}}g_Ig^z_{10}\mathcal{R}(\frac{q^\prime}{\alpha})^2 $\\
\hline
\end{tabular}
\end{table}
\end{center}
%\end{widetext}

\section{strong decays}\label{apt}

%\begin{widetext}
\begin{center}
\begin{table}[ht]
\caption{ The decay amplitudes for $|n \ ^{2S+1} L_{J}
\rangle\rightarrow |1^3S_1\rangle \mathbb{P}$. } \label{asb}
\begin{tabular}{|c|c|c|c|c|c|c }\hline\hline
$|n \ ^{2S+1} L_{J}\rangle$       &  $ J_z$                                                                    & amplitude\\
\hline
$1^1P_1  (1^+)    $ & $\pm 1$               &$ i g_Ig^+_{01}h\alpha  $\\
                    &  0             & $-i\frac{1}{\sqrt{2}}g_Ig^z_{01}\mathcal{R}\frac{q^\prime}{\alpha} -i\frac{1}{\sqrt{2}}g_Ig^z_{01}h\alpha  $\\
\hline
$1^3P_0  (0^+)    $ &                & forbidden\\
\hline
$1^3P_1   (1^+)   $ &$\pm 1$                &$i\frac{1}{2}g_Ig^z_{11}\mathcal{R}\frac{q^\prime}{\alpha} +i\frac{1}{2}g_I(g^z_{11}+\sqrt{2}g^+_{11})h\alpha  $\\
                    &0& $\sqrt{2}g_Ig^+_{11}h\alpha  $\\
\hline
$1^3P_2    (2^+)  $ & $\pm 1$                &$-i\frac{1}{2}g_Ig^z_{11}\mathcal{R}\frac{q^\prime}{\alpha} $\\
                    &0&0\\
\hline
$2^1S_0    (0^-)  $ & 0               &$\frac{1}{\sqrt{24}}g_Ig^z_{10}\mathcal{R}(\frac{q^\prime}{\alpha})^2 +\sqrt{\frac{1}{6}}g_Ig^z_{10}hq^\prime  $ \\
\hline
$2^3S_1     (1^-) $ & $\pm 1$               &$\pm\left\{\frac{1}{\sqrt{24}}g_Ig^z_{11}\mathcal{R}(\frac{q^\prime}{\alpha})^2F+\sqrt{\frac{1}{6}}g_Ig^z_{11}hq^\prime \right\}$ \\
                       &0& 0\\
\hline
$1^1D_2      (2^-)$  &  $\pm 1$               &$\frac{1}{\sqrt{2}}g^+_{01}g_Ihq^\prime  $\\
                     &0                      &$-\sqrt{\frac{1}{12}}g_Ig^z_{01}\mathcal{R}(\frac{q^\prime}{\alpha})^2 -\sqrt{\frac{1}{3}}g_Ig^+_{01}hq^\prime  $\\
\hline
$1^3D_1      (1^-)$  & $\pm 1$                &$\mp\left[\sqrt{\frac{1}{120}}g_Ig^z_{11}\mathcal{R}(\frac{q^\prime}{\alpha})^2 +\sqrt{\frac{5}{12}}g_Ig^+_{11}hq^\prime  \right]$\\
                       &0&0\\
\hline
$1^3D_2      (2^-)$  & $\pm 1$                &$\sqrt{\frac{1}{24}}g_Ig^z_{10}\mathcal{R}(\frac{q^\prime}{\alpha})^2 +\sqrt{\frac{3}{4}}g_Ig^+_{11}hq^\prime  $\\
                       &0& $ g^+_{11}g_Ihq^\prime  $\\
\hline
$1^3D_3      (3^-)$  & $\pm 1$                & $\mp \sqrt{\frac{1}{30}}g_Ig^z_{11}\mathcal{R}(\frac{q^\prime}{\alpha})^2  $\\
                      &0&0\\
\hline
\end{tabular}
\end{table}
\end{center}
%\end{widetext}

%\begin{widetext}
\begin{center}
\begin{table}[ht]
\caption{ The decay amplitudes for $|n \ ^{2S+1} L_{J}
\rangle\rightarrow |1^3P_0 \rangle\mathbb{P}$, where we have defined
$\mathcal{W}\equiv\mathcal{G}q(-1+\frac{q^{\prime 2}}{4\alpha^2})$,
$\mathcal{S}\equiv h\alpha(1-\frac{q^{\prime 2}}{2\alpha^2})$}
\label{asc}
\begin{tabular}{|c|c|c|c|c|c|c }\hline\hline
$|n \ ^{2S+1} L_{J}\rangle$       &  $ J_z$                                                                    & amplitude\\
\hline
$2^1S_0    (0^-)  $ & 0               &$i\frac{1}{3}g_Ig^z_{01}\mathcal{W}\frac{q^\prime}{\alpha}-i\frac{1}{3}g_Ig^z_{01}h\alpha \mathcal{A}  $ \\
\hline
$2^3S_1     (1^-) $ &               & forbidden \\
\hline
                     &0                      &$i\frac{\sqrt{2}}{3}g_Ig^z_{01}\mathcal{W}\frac{q^\prime}{\alpha}+i\frac{\sqrt{2}}{3}g_Ig^z_{01}h\alpha \mathcal{A}  $\\

$1^1D_2      (2^-)$  &  $\pm 1$               &0\\

                     &  $\pm 2$               &$-i\frac{\sqrt{2}}{3}g_Ig^+_{01}h\alpha  $\\
\hline

$1^3D_1      (1^-)$                       & & forbidden \\
\hline
                       &0& $-i \frac{\sqrt{6}}{3}g^+_{11}g_I  \mathcal{S}  $\\
    $1^3D_2      (2^-)$                   &$\pm 1$& 0\\
  &     $\pm 2$                & $-i \frac{2}{3}g^+_{11}g_I h\alpha F-i \frac{\sqrt{2}}{6}g^z_{11}g_I\mathcal{S} $\\

\hline
$1^3D_3      (3^-)$  & $\pm 2$                & $-i \frac{\sqrt{2}}{3}g^+_{11}g_Ih\alpha F+i \frac{1}{3}g^z_{11}g_I \mathcal{S} $\\
                      &0& 0\\
\hline
\end{tabular}
\end{table}
\end{center}
%\end{widetext}

%\begin{widetext}
\begin{center}
\begin{table}[ht]
\caption{ The decay amplitudes for $|n \ ^{2S+1} L_{J}
\rangle\rightarrow |1^1P_1 \rangle\mathbb{P}$.  } \label{asd}
\begin{tabular}{|c|c|c|c|c|c|c }\hline\hline
$|n \ ^{2S+1} L_{J}\rangle$      &  $(J^f_z, J^i_z)$                                                                    & amplitude\\
\hline
$2^1S_0    (0^-)  $ &                &forbidden \\
\hline
$2^3S_1     (1^-) $ &  $\pm(1,-1)$             & $-i\sqrt{\frac{2}{3}}g_Ig^+_{10}h\alpha (1+\frac{q^{\prime 2}}{4\alpha^2}) $ \\
                    &  $(0,0)$             & $-i\frac{1}{\sqrt{3}}g_Ig^z_{10}\mathcal{W}\frac{q^\prime}{\alpha} +i\frac{1}{\sqrt{3}}g_Ig^z_{10}h\alpha \mathcal{A}  $ \\
\hline

$1^1D_2      (2^-)$  &                 &forbidden\\

\hline

$1^3D_1      (1^-)$                       & $\pm(1,-1)$&$-i \sqrt{\frac{3}{20}}g_Ig^z_{10}\mathcal{G}q\frac{q^\prime}{\alpha} -i\sqrt{\frac{1}{30}}g_Ig^+_{10}\mathcal{S} $\\
                    &  $(0,0)$             & $-i\sqrt{\frac{4}{15}}g_Ig^z_{10}\mathcal{W}\frac{q^\prime}{\alpha} +i\sqrt{\frac{4}{15}}g_Ig^z_{10}h\alpha \mathcal{A}  $ \\
\hline \hline
%                       &(0,0)& \\
    $1^3D_2      (2^-)$                   &$\pm (1,-1)$& $-i\frac{1}{\sqrt{12}}g_Ig^z_{10}\mathcal{R}\frac{q^\prime}{\alpha} +i\frac{1}{\sqrt{24}}g_Ig^+_{10}h q^\prime\frac{q^\prime}{\alpha}  $\\
%  &     $\pm (2,-2)$                & 0\\

\hline
$1^3D_3      (3^-)$  & $\pm(1,-1)$                & $-i \sqrt{\frac{1}{30}}(\sqrt{2}g^z_{10}-g^+_{10})g_Ihq^\prime\frac{q^\prime}{\alpha}  $\\

                      &(0,0)& $i\sqrt{\frac{1}{5}}g_I[g^z_{10}\mathcal{W}\frac{q^\prime}{\alpha}-\sqrt{2}g^z_{10}h\alpha \mathcal{A} +2g^+_{10}\mathcal{S}] $ \\
\hline
\end{tabular}
\end{table}
\end{center}
%\end{widetext}

%\begin{widetext}
\begin{center}
\begin{table}[ht]
\caption{ The decay amplitudes for $|n \ ^{2S+1} L_{J}
\rangle\rightarrow |1^3P_1 \rangle\mathbb{P}$. } \label{ase}
\begin{tabular}{|c|c|c|c|c|c|c }\hline\hline
$|n \ ^{2S+1} L_{J}\rangle$       &  $(J^f_z, J^i_z)$                                                                    & amplitude\\
\hline
$2^1S_0    (0^-)  $ &                &forbidden \\
\hline
$2^3S_1     (1^-) $ &  $\pm(1,-1)$             & $-i\frac{1}{\sqrt{3}}g_Ig^+_{11}h\alpha (1+\frac{q^{\prime 2}}{4\alpha^2}) $ \\
                    &  $\pm(1,1)$             & $i\frac{1}{\sqrt{6}}g_Ig^z_{11}\mathcal{W}\frac{q^\prime}{\alpha} -i\frac{1}{\sqrt{6}}g_Ig^z_{11}h\alpha \mathcal{A}  $ \\
\hline

  &  $\pm(0, 2)$               & $\pm ig^+_{01}h \alpha  $\\
$1^1D_2      (2^-)$  &  $\pm(1,1)$               & $\pm i\frac{1}{\sqrt{2}}g^+_{01} \mathcal{S} $\\

&  $\pm(1,-1)$                 & $\pm i\frac{1}{2}g_Ig^z_{01}(\mathcal{R}\frac{q^\prime}{\alpha}+h \alpha)   $\\
\hline

$1^3D_1      (1^-)$                       & $\pm(1,-1)$&$-i\sqrt{\frac{1}{60}}g_Ig^+_{11}h\alpha(1+\frac{q^{\prime 2}}{2\alpha^2}) $\\

                    &  $\pm(1,1)$             & $-i\sqrt{\frac{1}{30}}g_Ig^z_{11}(\mathcal{W}\frac{q^\prime}{\alpha}-h\alpha \mathcal{A} -\frac{3}{2}\mathcal{S}) $ \\
\hline \hline

                       &$\pm(0,2)$                     & $\pm i\frac{1}{\sqrt{12}}g_Ig^z_{11}(\mathcal{R}\frac{q^\prime}{\alpha} +3 h \alpha ) $\\
    $1^3D_2      (2^-)$                   &$\pm (1,-1)$& $\pm[-i\sqrt{\frac{1}{3}}g_Ig^z_{11}h\alpha   -i\sqrt{\frac{1}{12}}g^+_{11}\mathcal{S} ]$\\

  &     $\pm (1,1)$                &  $\pm i\sqrt{\frac{1}{6}}g_Ig^z_{11}(\mathcal{W}\frac{q^\prime}{\alpha}-h\alpha \mathcal{A} -\frac{1}{2}\mathcal{S}) $ \\
\hline
                       &$\pm(0,2)$                     & $i\frac{1}{\sqrt{6}}g_Ig^z_{11}\mathcal{R}\frac{q^\prime}{\alpha} $\\
    $1^3D_3      (3^-)$                   &$\pm (1,-1)$& $-i\sqrt{\frac{1}{60}}g_Ig^+_{11}h q^\prime \frac{q^\prime}{\alpha}  $\\

  &     $\pm (1,1)$                &  $-i\sqrt{\frac{2}{15}}g_Ig^z_{11}[\mathcal{W}\frac{q^\prime}{\alpha}-h\alpha \mathcal{A} +\mathcal{S}] $ \\
\hline
\end{tabular}
\end{table}
\end{center}
%\end{widetext}

For a heavy-light meson $\bar{Q}q$, because the pseudoscalar mesons
$\mathbb{P}$ only couple with the light quarks, the strong decay
amplitudes for the process $\mathbb{M}_i \rightarrow \mathbb{M}_f
\mathbb{P} $ can be written as
\begin{eqnarray} \label{am}
&&\mathcal{M}(\mathbb{M}_i \rightarrow \mathbb{M}_f
\mathbb{P})\nonumber\\
& =& \left\langle \mathbb{M}_f\left|\left\{\mathcal{G}\vsig_1\cdot
 \textbf{q}
+h \vsig_1\cdot \textbf{p}'_1\right\}I_1 e^{-i\textbf{q}\cdot
\textbf{r}_1}\right|\mathbb{M}_i\right\rangle ,
\end{eqnarray}
with
\begin{eqnarray}
\mathcal{G}\equiv-\left(\frac{\omega_m}{E_f+M_f}+1\right),\ \
h\equiv\frac{\omega_m}{2\mu_q}.
\end{eqnarray}
$\mathbb{M}_i$  and $\mathbb{M}_f$ are the initial and final meson
wave functions, and they are listed in Tab. \ref{wfS}.
%$\textbf{p}'_1=-i\mathbf{\nabla}_1$ acts on the wave functions.
In the initial-meson-rest frame the energies and momenta of the
initial mesons $\mathbb{M}_i$ are denoted by $(E_i,
\textbf{P}_i$), while those of the final state mesons
$\mathbb{M}_f$ and the emitted pseudoscalar mesons $\mathbb{P}$
are denoted by $(E_f, \textbf{P}_f)$ and $(\omega_m, \textbf{q})$.
Note that $\textbf{P}_i=0$ and $\textbf{P}_f=-\textbf{q}$.

The form of Eq.(\ref{am}) is similar to that of in Refs.
\cite{Godfrey:1985xj,Godfrey:1986wj}, except that the factors
$\mathcal{G}$ and $h$ in this work have explicit dependence on the
energies of final hadrons. In the calculations, we select
$\mathbf{q}=q\hat{z}$, namely the meson moves along the $z$ axial.
Finally, we can work out the decay amplitudes for various process,
$\mathbb{M}\rightarrow |1^1S_0\rangle \mathbb{P}$,
$\mathbb{M}\rightarrow |1^3S_1\rangle \mathbb{P}$,
$\mathbb{M}\rightarrow |1^3P_0\rangle \mathbb{P}$,
$\mathbb{M}\rightarrow |1^1P_1\rangle \mathbb{P}$ and
$\mathbb{M}\rightarrow |1^3P_1\rangle \mathbb{P}$, which are
listed in Tabs. \ref{asa}--\ref{ase}, respectively.

Some analytical features can be learned here. From Tab. \ref{asa},
it shows that the decays of $1^1P_1$, $1^3P_1$, $2^1S_0$, $1^1D_2$
and $1^3D_2$ into $|1^1S_0\rangle \mathbb{P}$ are forbidden by
parity conservation. The decay amplitudes for $2^3S_1$, $2^3P_2$,
and $1^3D_3 \rightarrow |1^1S_0\rangle \mathbb{P}$ are
proportional to $\mathcal{R}$ (i.e. proportional to $q$),
$\mathcal{R} q^\prime/\alpha$ and $\mathcal{R} (q/\alpha)^2$,
respectively. This is crucial for understanding the small
branching ratios for $D^*(2007)\to D\pi$ as we will see later.

In contrast, the decay amplitude for $1^3P_0\rightarrow
|1^1S_0\rangle \mathbb{P}$ has two terms. One is proportional to
$\mathcal{R} q^\prime/\alpha$, while the other is proportional to
$\alpha$. Similarly, the decay amplitude for $2^3S_1\rightarrow
|1^1S_0\rangle \mathbb{P}$ and $2^3D_1\rightarrow |1^1S_0\rangle
\mathbb{P}$ also have two terms of which one is proportional to
$\mathcal{R} (q^\prime/\alpha)^2$, and the other to $q^\prime$.
This feature will have certain implications of their branching
ratio rates into different $|1^1S_0\rangle \mathbb{P}$ states.

From Tab. \ref{asb}, it shows that decays of $1^3P_0$ into
$|1^1S_0\rangle \mathbb{P}$ are forbidden. Among those three
helicity amplitudes $\mathcal{M}_{\pm}$ and $\mathcal{M}_{0}$, the
longitudinal one $\mathcal{M}_0$ vanishes for $1^3P_2$, $2^3S_1$,
$1^3D_1$, and $1^3D_3$ into $|1^1S_0\rangle \mathbb{P}$.
%From the table we can clearly see the structures of the helicity
%amplitudes.

From Tabs. \ref{asc}--\ref{ase}, we can see that the decays of
$2^3S_1$ and $1^3D_1$ into $|1^3P_0\rangle \mathbb{P}$, $2^1S_0$
and $1^1D_2$ into $|1^1P_1\rangle \mathbb{P}$, and $2^1S_0$ into
$|1^1P_1\rangle \mathbb{P}$ are forbidden parity conservation.
These selection rules are useful for the state classifications.

\section{calculations and analysis}\label{cd}

With the transition amplitudes, one can calculate the partial
decay width with
\begin{equation}\label{dww}
\Gamma=\left(\frac{\delta}{f_m}\right)^2\frac{(E_f+M_f)|\textbf{q}|}{4\pi
M_i(2J_i+1)} \sum_{J_{iz},J_{fz}}|\mathcal{M}_{J_{iz},J_{fz}}|^2 ,
\end{equation}
where $J_{iz}$ and $J_{fz}$ stand for the third components of the
total angular momenta of the initial and final heavy-light mesons,
respectively. $\delta$ as a global parameter accounts for the
strength of the quark-meson couplings. In the heavy-light meson
transitions, the flavor symmetry does not hold any more. Treating
the light pseudoscalar meson as a chiral field while treating the
heavy-light mesons as constitute quark system is an approximation.
This will bring uncertainties to coupling vertices and form
factors. Parameter $\delta$ is introduced to take into account
such an effect. It has been determined in our previous study of
the strong decays of the charmed baryons \cite{Zhong:2007gp}.
Here, we fix its value the same as that in
Ref.~\cite{Zhong:2007gp}, i.e. $\delta=0.557$.

In the calculation, the standard parameters in the quark model are
adopted. For the $u$, $d$, $s$, $c$ and $b$ constituent quark
masses we set $m_u=m_d=350$ MeV, $m_s=550$ MeV, $m_c=1700$ MeV and
$m_b=5100$ MeV, respectively. The decay constants for $\pi$, $K$,
$\eta$ and $D$ mesons, $f_\pi=132$ MeV, $f_K=f_{\eta}=160$ MeV,
and $f_D=226$ MeV, are used. For the masses of all the heavy-light
mesons the PDG values are adopted in the calculations \cite{PDG}.

To partly remedy the inadequate of the non-relativistic wave
function as the relative momentum $q$ increases, a commonly used
Lorentz boost factor is introduced into the decay
amplitudes~\cite{qkk,qk5,Zhong:2007fx},
\begin{eqnarray}
\mathcal{M}(q)\rightarrow \gamma_f \mathcal{M}(\gamma_fq),
\end{eqnarray}
where $\gamma_f=M_f/E_f$. In most decays, the three momentum carried
by the final state mesons are relatively small, which means the
non-relativistic prescription is reasonable and corrections from
Lorentz boost are not drastic.

\begin{table}[ht]
\caption{Predictions of the strong decay widths (in MeV) for the
heavy-light mesons. For comparison, the experimental data and some
other model predictions are listed.} \label{width}
\begin{tabular}{|c|c|c|c|c|c|c|c|c|c|c }\hline\hline
    & notation    &  channel  &  $\Gamma( \mathrm{this\ work} ) $ & $\Gamma$\cite{Close:2005se}& $\Gamma$\cite{Godfrey:2005ww}   & $\Gamma $ \cite{Orsland:1998de}  & $\Gamma$\cite{Zhu:1998wy,Dai:1998ve} & $\Gamma$\cite{Falk:1995th}    & $\Gamma_{exp}$   \\
\hline \hline
     $D^*(2007)^0$    &$(1^3S_1) 1^-$    &  $D^0\pi^0$   &  58 keV    &  16 keV  &         &  39 keV &          &                      &  $<2.1$ MeV                        \\
     $D^*(2010)^+$    &                  &  $D^0\pi^+$   &  77 keV    &  25 keV  &         &  60 keV &          &                      &  $64\pm 15$ keV                    \\
                      &                  &  $D^+\pi^0$   &  35 keV    &  11 keV  &         &  27 keV &          &                      &    $29\pm7$  keV                      \\
\hline
     $D^*_0(2352)$    &$(1^3P_0) 0^+$    &  $D\pi$       &  248       &          &  277    &         &          &                   &  $261\pm50$                     \\
     $D^*_0(2403)$    &                  &               &  266       &  283     &         &         &          &                      &  $283\pm58$                      \\
\hline
     $D_1(2420)$      &$(1^1P_1)  1^+$   &  $D^*\pi$     &   84       &          &         &         &          &                      &                       \\
                      & $(P_1) 1^+$      &               &   21.6     &   22     &   25    &         &          &                  &  $25\pm 6$                        \\
\hline
     $D'_1(2430)$     &$(1^3P_1)  1^+$   &  $D^*\pi$     &   152      &          &         &         &          &                      &                        \\
                      &$( P_1')1^+$      &               &   220      &   272    &  244    &         &          &                      &  $384\pm117$                       \\
\hline
 $D^*_2(2460)^0$      &$(1^3P_2) 2^+$    &  $D\pi$       & 39         & 35       & 37      &         & 13.7     &                      &                        \\
                      &                  & $D^*\pi$      & 19         & 20       & 18      &         & 6.1      &                      &                        \\
                      &                  & $D \eta$      & 0.1        & 0.08     &         &         &          &                      &                        \\
                      &                  & total         & 59         & 55       & 55      &         & 20       &                      &     $43\pm 4$                    \\
 \hline\hline
 $D_{s1}(2536)$       &$(1^1P_1)   1^+$  &  $D^*K$       & 59         &          &         &         &          &                      &                        \\
                      &$  (P_1) 1^+$     &               & 0.35       & 0.8      & 0.34    &         &          &                      &        $< 2.3$             \\
\hline
%$D'_{s1}(2460)$      &$  (P_1') 1^+$    &  $D^*K$      &  0          &          &         &         &          &                      &      $< 5.5$                  \\
%
%\hline
 $D^*_{s2}(2573)$     &$(1^3P_2) 2^+$    &  $DK$        & 16          & 27       & 20      &         &          &                       &                        \\
                      &                  & $D^*K$       & 1           & 3.1      & 1       &         &          &                       &                        \\
                      &                  & $D_s \eta$   & 0.4         & 0.2      &         &         &          &                       &                        \\
                      &                  & total        & 17          & 30       & 21      &         &          &                       &     $15^{+5}_{-4}$                   \\
 \hline
$D^*_{sJ}(2860)$      &$(1^3D_3) 3^-$    & $DK$         & 27          &          &         &         &          &                       &                        \\
                      &                  & $D^*K$       & 11          &          &         &         &          &                       &                        \\
                      &                  & $D_s \eta$   & 3           &          &         &         &          &                       &                        \\
                      &                  & $D^*_s \eta$ & 0.3         &          &         &         &          &                       &                        \\
                      &                  & $D K^*$      & 0.4         &          &         &         &          &                       &                        \\

                      &                  & total        & 42          &          &         &         &          &                       &     $48\pm 17$                   \\
 \hline\hline
$B^*_0(5730)$         &$  (^3P_0) 0^+$   &  $B\pi$      &  $272$        &          &         & 141       &   250    &                          &                     \\

\hline
$B_1(5725)$           &$  (P_1) 1^+$     &  $B^*\pi$    &  30       &          &         &  20       &          &                       &      $20\pm 12$                 \\

\hline
                      &$(1^3P_1)  1^+$   &  $B^*\pi$    &   153       &          &         &           &          &                       &                        \\
$B'_1(5732)$          &$  (P_1') 1^+$    &  $B^*\pi$    &   219        &         &         & 139       &   250    &                       &      $128\pm 18$                 \\

\hline
$B^*_2(5740)$         &$(1^3P_2) 2^+$    &  $B \pi$     & 25          &          &         & 15        &   3.9    &                       &                        \\
                      &                  & $B^*\pi$     & 22         &          &         & 14        &   3.4    &                       &                        \\
                      &                  & total        & 47          &          &         & 29        &   7.3    &   $16\pm 6$                     &     $22^{+7}_{-6}$                   \\
 \hline\hline
$B^*_{s0}(5800)$      & $  (^3P_0) 0^+$  &  $B K$       & 227      &          &         &           &          &                       &                        \\

 \hline
$B_{s1}(5830)$        & $  (P_1) 1^+$    &  $B^* K$     & $0.4\sim 1$   &          &         &           &          &   $3\pm 1$                  &         1               \\

 \hline
$B_{s1}'(5830)$       & $  (P_1') 1^+$   &  $B^* K$     & $149$         &          &         &           &          &                       &                        \\

 \hline
$B^*_{s2}(5839)$      &$(1^3P_2) 2^+$    &  $B K$       & 2           &          &         &           &          &                       &                        \\
                      &                  & $B^*K$       & $0.12$      &          &         &           &          &                       &                        \\
                      &                  & total        & 2           &          &         &           &    &   $7\pm 3$                     &      1                 \\
 \hline\hline

\end{tabular}
\end{table}

\subsection{Strong decays of 1$S$ states}

Due to isospin violation, $D^{*+}$ is about 3 MeV heavier than the
neutral $D^{*0}$~\cite{PDG}. This small difference leads to a
kinematic forbiddance of $D^{*0}\to D^+\pi^-$, while $D^{*+}\to
D^0\pi^+$ and $D^+\pi^0$ are allowed, but with a strong kinematic
suppression. Nevertheless, it shows by Tab. \ref{asa} that the
decay amplitudes of $1S$ states are proportional to the final
state momentum $q$. For the decays of $D^{*0}\to D^0\pi^0$,
$D^{*+}\to D^0\pi^+$ and $D^+\pi^0$ of which the decay thresholds
are close to the $D^*$ masses, it leads to further dynamic
suppressions to the partial decay widths. As shown in Tab.
\ref{width}, our calculations are in remarkable agreement with the
experimental data. Since $q$ is small, the form factor corrections
from quark model are negligibly small. One would expect that the
ratio $\Gamma(D^0\pi^+)/\Gamma(D^+\pi^0)\simeq 2$ is then
dominated by the isospin factor $g_I$, which agrees well with the
prediction in Ref.~\cite{Close:2005se}.

\subsection{Strong decays of 1$P$ states}

In the $LS$ coupling scheme, there are four 1$P$ states: $^3P_0,
^3P_1,^3P_2$ and $^1P_1$. For $^3P_0$, its transition to
$|1^3S_1\rangle \mathbb{P}$ is forbidden. States of $^1P_1$, and
$^3P_1$ can couple into $|1^3S_1\rangle \mathbb{P}$, but not
$|1^1S_0\rangle \mathbb{P}$. In contrast, $^3P_2$ can be coupled
to both $|1^3S_1\rangle \mathbb{P}$ and $|1^1S_0\rangle
\mathbb{P}$. In the decay amplitudes of $^3P_0$, $^1P_1$, and
$^3P_1$, the term $h\alpha F$ dominates the partial decay widths,
and usually their decay widths are much broader than that of
$^3P_2$. Between the amplitudes of the $ ^1P_1$ and $^3P_1$
decays, we approximately have:
\begin{eqnarray}
\mathcal{M}(^1P_1\rightarrow |1^3S_1\rangle \mathbb{P})_{J_z}\simeq
\frac{1}{\sqrt{2}}\mathcal{M}(^3P_1\rightarrow |1^3S_1\rangle
\mathbb{P})_{J_z},
\end{eqnarray}
since the term $\mathcal{R}\frac{q^\prime}{\alpha}F$ is negligible
when the decay channel threshold is close to the initial meson
mass. As a consequence, the decay widths of the $ ^1P_1$ states
are narrower than those of $^3P_1$.

\subsubsection{$^3P_0$ states}

$D^*_0(2400)$ is listed in PDG~\cite{PDG} as a broad $^3P_0$
state. Its mass values from Belle~\cite{belle-04d} and FOCUS
Collaboration~\cite{focus-04a} are quite different though the
FOCUS result is consistent with the potential quark model
prediction of 2403 MeV~\cite{Godfrey:1985xj}. In experiment, only
the $D\pi$ channel are observed since the other channels are
forbidden. The term $h \alpha F$ in the amplitude, which is in
proportion to the oscillator parameter $\alpha$, accounts for the
broad decay width. By applying the PDG averaged mass 2352 MeV and
the FOCUS value 2403 MeV, its partial decay widths into $D\pi$ are
calculated and presented in Tab. \ref{width}. They are in good
agreement with the data~\cite{PDG,focus-04a}.

In the $B$ meson sector, $B^*_0$ and $B^*_{s0}$, as the $^3P_0$
states, have not been confirmed in any experiments. The predicted
mass for $B^*_0$ and $B^*_{s0}$ mesons are about 5730 MeV and 5800
MeV, respectively \cite{Vijande:2007ke}. Their strong decays only
open to $B\pi$ or $BK$. Applying the theory-predicted masses, we
obtain broad decay widths for both states, i.e.
$\Gamma(B^*_0)=272$ MeV, and $\Gamma(B^*_{s0})=227$ MeV,
respectively. Our prediction of $\Gamma(B^*_0)=272$ MeV is
compatible with the QCDSR prediction $\Gamma(B^*_0)\simeq 250$ MeV
\cite{Zhu:1998wy}.  Such broad widths may explain why they have
not yet been identified in experiment.

\subsubsection{$^3P_2$ states}

In PDG, the decay width of $D^{*}_2(2460)^0$ is $\Gamma=43\pm 4$
MeV and that of $D^{*}_2(2460)^{\pm}$ is $\Gamma=29\pm 5$ MeV.
Since there is no obvious dynamic reason for such a significant
difference, it may simply be due to experimental uncertainties.
Our prediction $\Gamma=59$ MeV as a sum of the partial widths of
$D\pi$, $D^*\pi$ and $D\eta$, is comparable with the data.
Nevertheless, the partial width ratio
\begin{eqnarray}
R\equiv\frac{\Gamma(D\pi)}{\Gamma(D^*\pi)}\simeq 2.1
\end{eqnarray}
obtained here is also in good agreement with the data $R\simeq
2.3\pm 0.6$~\cite{PDG}.

$D^*_{s2}(2573)$ is assigned to be a $^3P_2$ state. Its total
width is $\Gamma_{exp}=15^{+5}_{-4}$ and the width ratio between
$D^*K$ and $DK$ is
$R\equiv\Gamma(D^*K)/\Gamma(DK)<0.33$~\cite{PDG}. Our predictions
for the total width and ratio $R$ are
\begin{eqnarray}
\Gamma &=&17 \ \mathrm{MeV}, \\
 R&\equiv&\frac{\Gamma(D^*K)}{\Gamma(D K)}\simeq 6\%,
\end{eqnarray}
which are consistent with the data.

Notice that the width of $D^*K$, $\sim 1$ MeV, is
one-order-of-magnitude smaller than that of $DK$. Apart from the
kinematic phase space suppression, its transition amplitude also
suffers dynamic suppressions since it is proportional to
$\mathcal{R}{q^\prime}/\alpha$. This explains its absence in
experiment. Although the decay channel $D\eta /D_s \eta$ is also
opened for $D^*_2(2460) /D^*_{s2}(2573)$, its partial width is
negligibly small, i.e. $< 1$ MeV.

In the $B$ meson sector, a candidate of $^3P_2$ state is from CDF
collaboration with mass \cite{B:2007CDF}.%[****** CITE CDF PAPER]
\begin{eqnarray}
M(B^*_2)=5740\pm 2\pm 1 \ \mathrm{MeV}.
\end{eqnarray}
D0 collaboration also observed the same state with slightly
different masses, $M(B^*_2)=5746.8\pm 2.4\pm 1.7 \ \mathrm{MeV}$
\cite{Abazov:2007vq}. By assigning $B^*_2$ as a $^3P_2$ state, the
predicted total width as sum of $B\pi$ and $B^*\pi$ is
\begin{eqnarray}
\Gamma(B^*_2)\simeq 47\  \mathrm{MeV},
\end{eqnarray}
which is consistent with the CDF measurement
$\Gamma(B^*_2)_{exp}\simeq 22^{+7}_{-6}$ MeV. It shows that these
two partial widths of $B\pi$ and $B^*\pi$ are comparable with each
other, and the predicted width ratio is
\begin{eqnarray}
R\equiv\frac{\Gamma( B^*\pi)}{\Gamma( B^*\pi)+\Gamma(B\pi)}=0.47 \
.
\end{eqnarray}
This is also in good agreement with the recent D0 data $R=0.475\pm
0.095\pm 0.069$ \cite{Abazov:2007vq}. %[****** CITE D0 PAPER].

CDF collaboration also reported an observation of $B^*_2$'s strange
analogue $B^*_{s2}$ \cite{:2007tr}, of which the mass is
\begin{eqnarray}
M(B^*_{s2})=5840\pm  1 \ \mathrm{MeV}.
\end{eqnarray}
With this mass, we obtain its partial decay widths,
$\Gamma(B^*K)=0.12$ MeV and $\Gamma(BK)=2$ MeV, respectively. This
gives its strong decay width and width ratio between $B^*K$ and
$BK$:
\begin{eqnarray}
\Gamma(B^*_{s2})\simeq  2 \ \mathrm{MeV},\
R\equiv\frac{\Gamma(B^*K)}{\Gamma(BK)}\simeq 6\%.
\end{eqnarray}
The decay width is in good agreement with the data
$\Gamma(B^*_{s2})_{exp}\sim 1 $ MeV \cite{Akers:1994fz}. It also
shows that the partial width of $B^*K$ channel is negligible
small, and will evade from observations in experiment. But a
measurement of $\Gamma(BK)$ with improved statistics should be
very interesting.

\subsubsection{The mixed states}

The $D_{1}(2420)$ and $D'_1(2430)$ listed in PDG~\cite{PDG}
correspond to a narrow and broad state, respectively. Their two
body pionic decays are only seen in $D^*\pi$. If they are pure $P$
wave states, they should be correspondent to $ ^1P_1$ and $^3P_1$.
The calculated decay widths by assigning them as $^1P_1$ and
$^3P_1$, are listed in Tab. \ref{width}. It shows that $D_1(2420)$
as a pure $ ^1P_1$  state, its decay width is overestimated by
about an order, while $D'_1(2430)$ as a pure $ ^3P_1$  state, its
decay width is underestimated by about a factor of 2. Similarly
large discrepancies are also found if one simply exchanges the
assignments.  Thus, the pure $ ^1P_1$ and $^3P_1$ scenario cannot
explain the nature of $D_1(2420)$ and $D'_1(2430)$.

\begin{center}
\begin{figure}[ht]
\centering \epsfxsize=8 cm \epsfbox{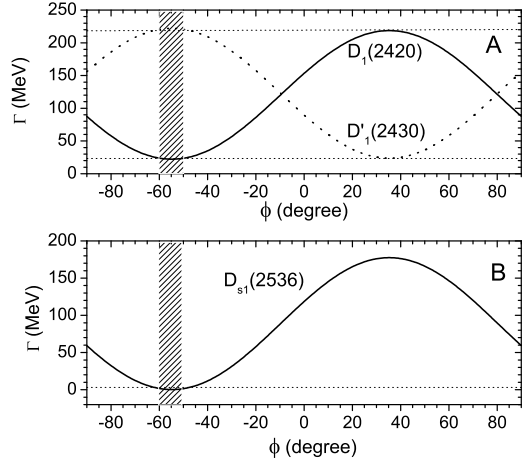} \caption{  The
decay widths of $D_1(2420)$, $D_1(2430)$ and $D_{s1}(2536)$ as
functions of the mixing angle $\phi$.} \label{mix}
\end{figure}
\end{center}

Since the heavy-light mesons are not charge conjugation
eigenstates, state mixing between spin $\textbf{S}=0$ and
$\textbf{S}=1$ states with the same $J^P$ can occur. The physical
states with $J^P=1^+$ would be given by
\begin{eqnarray}
|P_1'\rangle=+\cos (\phi) |^1P_1\rangle+\sin(\phi)|^3P_1\rangle, \\
|P_1\rangle=-\sin (\phi) |^1P_1\rangle+\cos(\phi)|^3P_1\rangle.
\end{eqnarray}
Our present knowledge about the $D_1(2420)$ and $D'_1(2430)$
mixing is still limited. The determination of the mixing angle is
correlated with quark potential, and masses of the
states~\cite{Godfrey:1986wj}. An analysis by
Ref.~\cite{Close:2005se} suggests that a mixed state dominated by
$S$-wave decay will have a broad width, and the $D$-wave-dominant
decay will have a narrow one. By assuming that the heavy quark
spin-orbit interaction is positive, this leads to an assignment of
$D'_1(2430)$ and $D_1(2420)$ as a mixed $|P_1'\rangle$ and
$|P_1\rangle$, respectively, with a negative mixing angle
$\phi=-54.7^\circ$. However, this will lead to that the mass of
$D_1$ is heavier that of the $D'_1$ for which the present
experimental precision seem unable to rule out such a
possibility~\cite{PDG}. An additional piece of information
supporting such a scenario is that a positive spin-orbit
interaction will lead to a heavier $2^+$ state than $0^+$ which
indeed agrees with experiment~\cite{Close:2005se}.

In our calculation, we plot the pionic decay widths of the mixed
states $|P_1'\rangle$ and $|P_1\rangle$ as functions of $\phi$ in
Fig. \ref{mix}. By looking for the best description of the
experimental data, we determine the optimal mixing angle. It shows
that with $\phi=-(55\pm 5)^\circ$, $D_1(2420)$, as the
$|P_1\rangle$ mixed state, has a narrow decay width of
$\Gamma\simeq 22$ MeV. This value agrees well with the
experimental data (see Tab. \ref{width}). Our prediction for the
width of $D'_1(2430)$ as a $|P'_1\rangle$ broad state is
$\Gamma\simeq 217$ MeV, which also agrees with the data
\cite{PDG}. Note that there are still large uncertainties with the
$D'_1(2430)$ measurements, and further experimental investigation
is needed.

Such a mixing scenario may occur within the $D_{s1}$ states, which
leads to $D_{s1}(2460)$ and $D_{s1}(2536)$ as the mixed
$|P'_1\rangle$ and $|P_1\rangle$, respectively. Note that
$D_{s1}(2460)$ has a relatively light mass which is below the $D^*K$
threshold, and also slightly below the $DK$ threshold. Therefore,
its strong decay is nearly forbidden, which makes it a narrow state.
On the other hand, $D_{s1}(2536)$, as a $|P_1\rangle$ mixed state
with the mixing angle $\phi=-(55\pm 5)^\circ$, can give a decay
width consistent with the data ($\Gamma< 2.3$ MeV)
\begin{equation}
\Gamma(D_{s1}(2536))\simeq  0.4\sim 2.5 \ \mathrm{MeV}.
\end{equation}
In contrast, if $D_{s1}(2536)$ is a pure $^1P_1$ state, its decay
width will be 59 MeV, which is overestimated by a factor of 20.

We also derive the width ratio
\begin{equation}
R\equiv\frac{\Gamma(D^*(2007)^0K^+)}{\Gamma(D^*(2010)^+K^0)}\simeq
1.2\sim 1.7,
\end{equation}
which is consistent with the experimental result, $R=1.27\pm
0.27$. In Fig.~\ref{mix}(B), the change of the strong decay width
$\Gamma(D_{s1}(2536))$ in terms of the mixing angle $\phi$ is
presented by treating it as mixed $|P_1\rangle$ state. It should
be mentioned that the recent measurements of the angular
decomposition of $D_{s1}(2536)^+\rightarrow D^{*+}K^0_S$ indicate
configuration mixings within $D_{s1}(2536)$ \cite{:2007dya}.

In the $B$ meson sector, two new narrow excited $B_1$ and $B_{s1}$
mesons are recently reported by CDF, with masses
\begin{eqnarray}
M(B_1)=5725\pm 2\pm 1 \ \mathrm{MeV},\\
M(B_{s1})=5829\pm 1 \ \mathrm{MeV}.
\end{eqnarray}
D0 collaboration also observed the same $B_1$ state with a
slightly different mass, $M(B_1)=5720\pm 2.4\pm 1.4 \
\mathrm{MeV}$.

The narrowness of these two axial vector states make them good
candidates as the narrow heavy partners in the state mixing. $B_1$
as a $|P_1\rangle$ state, its strong decay width to $B^*\pi$ is
predicted to be
\begin{eqnarray}
\Gamma(B_1)\simeq 30\  \mathrm{MeV}.
\end{eqnarray}
With the strong decay widths for $B_2^*\to B\pi$ and $B^*\pi$
calculated, we obtain the strong decay width ratio
\begin{eqnarray}
R\equiv\frac{\Gamma(B_1)}{\Gamma(B_1)+\Gamma(B^*_2)}=0.34,
\end{eqnarray}
which are in good agreement with the recent D0 data $R=0.477\pm
0.069\pm 0.062$ \cite{Abazov:2007vq}.%[****** CITE D0 PAPER].

Note that $B_J^*(5732)$ in PDG~\cite{PDG} is a broad state with
$\Gamma_{exp}=128\pm 18$ MeV.  The PDG averaged mass is $5698\pm
8$ MeV which makes it lighter than $B_1(5725)$. This makes it a
natural candidate as the mixed light partner $|P_1'\rangle$, for
which the predicted width is $\Gamma(B_1')=219$ MeV, this result
is compatible with the QCDSR prediction $\Gamma(B_1')\simeq 250$
MeV. As a test, we find that $B_J^*(5732)$ as a pure $^3P_1$ state
its decay width is 153 MeV, which seems to agree well with the PDG
suggested value. Whether $B_J^*(5732)$ is a mixed state
$|P'_1\rangle$, a pure $^3P_1$ state, or other configurations,
needs further improved experimental measurement.

Similarly, $B_{s1}$ as a $|P_1\rangle$ state, its strong decay
width and decay width ratio to the sum of $B_{s1}$ and $B_{2s}^*$
widths are
\begin{eqnarray}
&& \Gamma(B_{s1})\simeq  0.4\sim 1\  \mathrm{MeV},\\
&&
R\equiv\frac{\Gamma(B_{s1})}{\Gamma(B_{s1})+\Gamma(B^*_{2s})}=0.02\sim
0.6.
\end{eqnarray}
The predicted width $\Gamma(B_{s1})$ agrees with the data
$\Gamma(B_{s1})_{exp}\sim 1$ MeV \cite{Akers:1994fz}.

Since the mass of $|P_1'\rangle$ is slightly lower than that of
$|P_1\rangle$,  the mass of $B'_{s1}$ (as a $|P_1'\rangle$ state)
should be less than 5830 MeV. If we assume the mass of $B'_{s1}$
is around 5830 MeV, which gives a broad decay width to $B^*K$
channel
\begin{eqnarray}
&& \Gamma(B'_{s1})\simeq  149\  \mathrm{MeV}.
\end{eqnarray}

We should point out that the mass of $B'_{s1}$ is most likely
below the threshold of $B^*K$, thus, the decay $B'_{s1}\rightarrow
B^*K$ is kinematically forbidden. In this case the decay width of
$B'_{s1}$ will be very narrow. Its decay properties should be
similar to those of $D_s(2460)$. The isospin violation decay
$B'_{s1}\rightarrow B_s^*\pi$ and radiative decay
$B'_{s1}\rightarrow B_s^*\gamma$ will be the dominant decay modes.
A recent study of this scenario was given by Wang with light-cone
sum rules~\cite{Wang:2008wz}.

\subsection{Strong decays of 2$S$ states}

\begin{center}
\begin{figure}[ht]
\centering \epsfxsize=8 cm \epsfbox{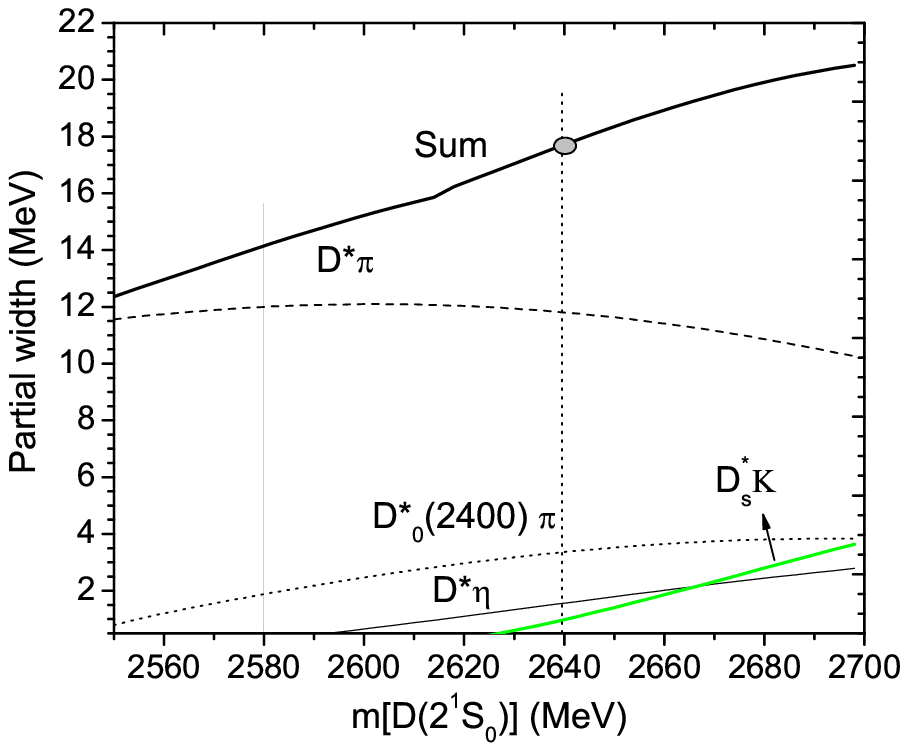} \centering
\epsfxsize=8 cm \epsfbox{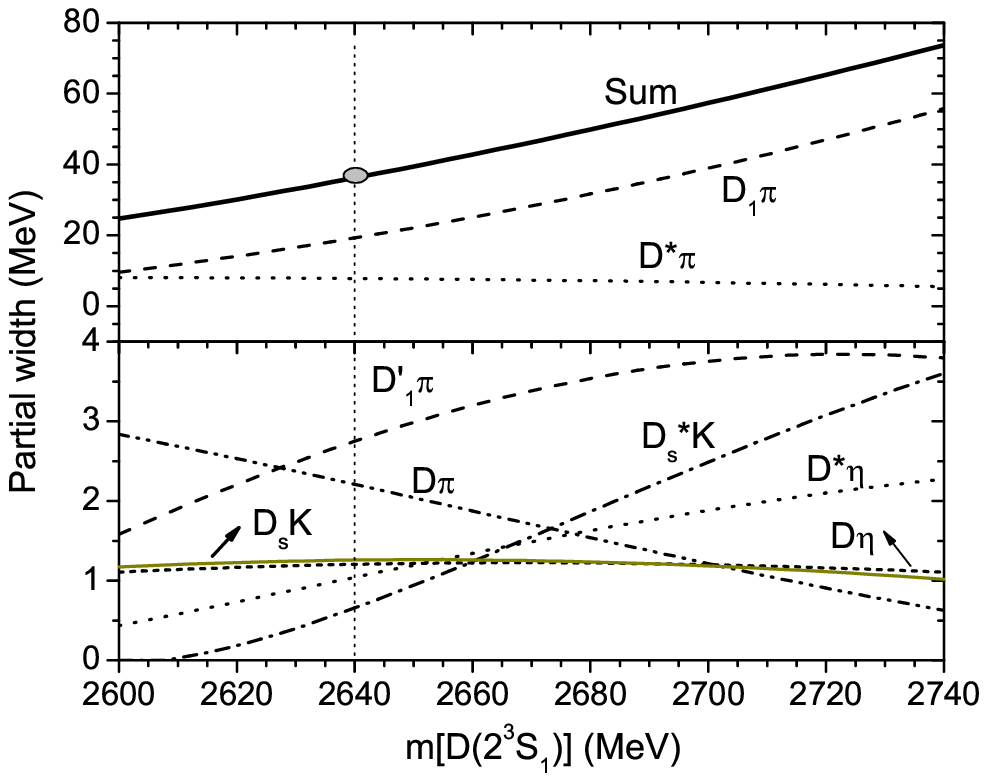}\caption{  The partial widths of
$D(2^1S_0)$ and $D(2^3S_1)$ as functions of the mass.}\label{2sd}
\end{figure}
\end{center}

\begin{center}
\begin{figure}[ht]
\centering \epsfxsize=8 cm \epsfbox{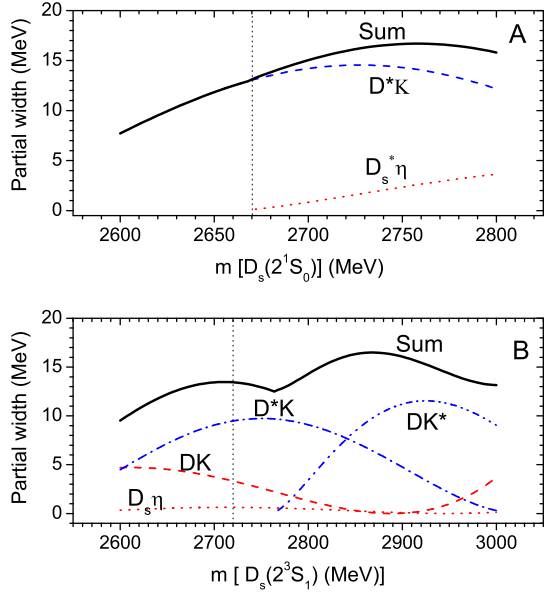} \caption{  The partial
widths of $D_s(2^1S_0)$ and $D_s(2^3S_1)$ as functions of the
mass.}\label{2sds}
\end{figure}
\end{center}

The radially excited heavy-light mesons are still not
well-established in experiment, although there are several
candidates,  such as $D^{*}(2640)^{\pm}$ \cite{Abreu:1998vk},
$D^*_{sJ}(2632)$ \cite{Evdokimov:2004iy} and $D^*_{sJ}(2700)$
\cite{Aubert:2006mh,:2007aa}. In theory, the radially excited $D$
states $2^1S_0$ and $2^3S_1$ were predicted to have masses $\sim
2.58$ and $\sim 2.64$ GeV, respectively \cite{Godfrey:1985xj},
while the radially excited $D_s$ states $2^1S_0$ and $2^3S_1$ were
$\sim 2.6$ and $\sim 2.7$ GeV, respectively
\cite{Godfrey:1985xj,Close:2006gr}. In this section, we study the
strong decays of these excited states into various channels. The
mass uncertainties bring uncertainties into the predicted partial
decay widths. Occasionally, some of the predicted partial widths
exhibit sensitivities to the meson masses. Therefore, we present
the strong decay widths of the $D$ and $D_s$ radially excited
states as functions of their masses within a reasonable range as
predicted by theory, and plot them along with their partial decay
widths in Figs. \ref{2sd} and \ref{2sds}, respectively. For a
given initial mass, by comparing
 the relative magnitudes among different partial
widths from theoretical prediction and experimental measurement,
one can extract additional information about the initial meson
quantum numbers.

\subsubsection{Radially excited $D$ mesons}

For a $2^1S_0$ state with a mass around 2.64 GeV, it can decay
into $D^*\pi$, $D^*\eta$, $D^*_sK$ and $D^*_0(2400)\pi$. In Fig.
\ref{2sd}, the partial widths and total strong decay width are
shown for a mass range. In these channels, the $D^*\pi$ dominates
the decays, the total decay width is $\sim 14$ MeV at
$m(D(2^1S_0))$=2.58 GeV, and it shows a flat behavior. Note that
the threshold of $D^*_s K$ channel is very close to 2.64 GeV. Some
sensitivities to this open channel thus occur in a mass range
around 2.6 GeV. It shows that this width increases quickly with
the masses and will compete again $D^*\eta$.

For the radially excited state $2^3S_1$, its dominant decay
channel is $D_1(2420)\pi$, while the other partial widths are much
smaller (see lower panel of Fig. \ref{2sd}). Again, the $D^*\pi$
partial width appears insensitive to the initial $D$ meson mass.

Comparing the decay patterns between $2^1S_0$ and $2^3S_1$ in Fig.
\ref{2sd}, we find it useful for clarifying $D^{*}(2640)^{\pm}$.
This state was first seen by BELPHI in $D^{*+}\pi^+\pi^-$ channel
with a narrow width $< 15$ MeV \cite{Abreu:1998vk}, but has not
yet been confirmed by other experiments. If it is a genuine
resonance, it will fit better into the $2^3S_1$ state instead of
$2^1S_0$ due to its dominant decays into $D^{*+}\pi^+\pi^-$ which
can occur via the main channel $D^{*}(2640)^+\to
D_1(2420)^0\pi^+$. In contrast, the assignment to a $2^1S_0$ state
will imply a dominant decay channel to $D^*\pi$ which is not
supported by the data. Although the predicted width $\sim 34$ MeV
overestimates the data by nearly a factor of two, it should be
more urgent to establish it in experiment and have more precise
measurement of its partial decay widths to both $D^*\pi$ and
$D^*\pi\pi$.

\subsubsection{Radially excited $D_s$ mesons}\label{radial-ds}

There are experimental signals for several excited $D_s$ states,
i.e. $D_{sJ}(2632)$~\cite{Evdokimov:2004iy}, $D_{sJ}(2690)$,
$D_{sJ}(2860)$~\cite{Aubert:2006mh}, and
$D_{sJ}(2708)$~\cite{:2007aa,belle-2006} for which the
spectroscopic classification is still unsettled. The
$D_{sJ}(2690)$ and $D_{sJ}(2708)$ are likely to the same state as
they have similar masses and both are broad. We shall compare
these experimental observations with our model predictions in
order to learn more about their spectroscopic classifications.

$D_{sJ}(2632)$ was reported by SELEX as a narrow state, i.e.
$\Gamma< 17$ MeV, in $D_s\eta$ and $DK$ channels
\cite{Evdokimov:2004iy}. The measured ratio of the partial widths
is
\begin{eqnarray}
R\equiv\frac{\Gamma(D^0 K^+ )}{ \Gamma(D_s \eta)}=0.16\pm 0.06.
\end{eqnarray}
Its dominant decay into $D_s\eta$ makes it difficult to assign it
into any simple $c\bar{q}$ scenario~\cite{barnes-004}. In
particular, since a $2^1S_0$ state is forbidden to decay into
$D_s\eta$ and $DK$, it rules out $D_{sJ}(2632)$ to be a radially
excited $0^-$.

As shown by Fig. \ref{2sds}, the decay of a $2^3S_1$ state turns
to be dominated by $D^*K$ and possibly $DK$, while its decay into
$D_s\eta$ is rather small. Therefore, a simple $2^3S_1$ cannot
explain its decay pattern as well. Some more investigations of the
nature of $D_{sJ}(2632)$ can be found in the literature, and here
we restrict our attention on the output of our model calculations.

$D^*_{sJ}(2860)$ and $D^*_{sJ}(2690)$ from
BaBar~\cite{Aubert:2006mh} (or $D_{sJ}(2708)$ from
Belle~\cite{:2007aa,belle-2006}) have widths
\begin{eqnarray}
\Gamma(D_{sJ}(2860))=48\pm 17 \ \mathrm{MeV},\\
\Gamma(D_{sJ}(2690))=112\pm 43 \ \mathrm{MeV} ,
\end{eqnarray}
and both are observed in the $DK$ channel, and no evidences are
seen in $D^*K$ and $D_s\eta$ modes. Compare these with Fig.
\ref{2sds}, it shows that neither of them can easily fit in
$2^1S_0$ or $2^3S_1$.

By fixing the masses of $2^1S_0$ and $2^3S_1$ states as suggested
by the quark model~\cite{Close:2006gr}, i.e. $m(D_s(2^1S_0))=2.64$
GeV and $m(D_s(2^3S_1))=2.71$ GeV, we obtain their strong decay
widths
\begin{eqnarray}
\Gamma(D_{s}(2^1S_0))\simeq 11 \ \mathrm{MeV},\\
\Gamma(D_{s}(2^3S_1))\simeq 14 \ \mathrm{MeV},
\end{eqnarray}
which turn out to be narrow.  For $D_s(2^1S_0)$, the predicted
dominant decay mode is $D^*K$, while the $DK$ channel is
forbidden. For $D_s(2^3S_1)$, there are two main decay channels
$D^*K$ and $DK$, and they give a ratio of
\begin{eqnarray}
R\equiv\frac{\Gamma(D^*K)}{\Gamma(DK)}\simeq 2.6 \ .
\end{eqnarray}
The $D_s \eta$ channel is also opened, but is negligibly small in
comparison with $DK$ and $D^*K$.

As $D^*_{sJ}(2860)$ has a relatively larger mass to fit in a
$D$-wave state, we shall examine it with $D$-wave decays in the
following subsection.

\begin{widetext}
\begin{center}
\begin{figure}[ht]
\centering \epsfxsize=12 cm \epsfbox{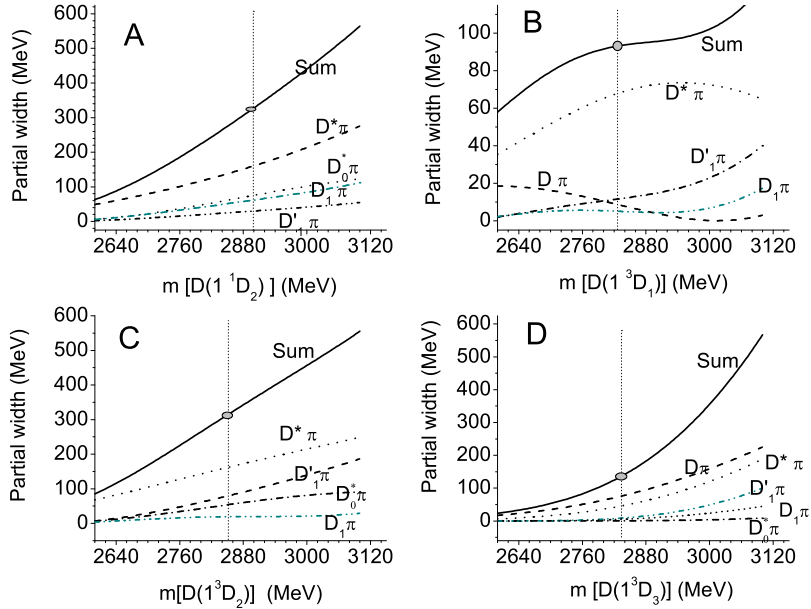} \caption{  The
partial widths of $D(1^1D_2)$, $D(1^3D_1)$, $D(1^3D_2)$ and
$D(1^3D_3)$ as functions of the mass.}\label{dw}
\end{figure}
\end{center}
\end{widetext}

\begin{widetext}
\begin{center}
\begin{figure}[ht]
\centering \epsfxsize=12 cm \epsfbox{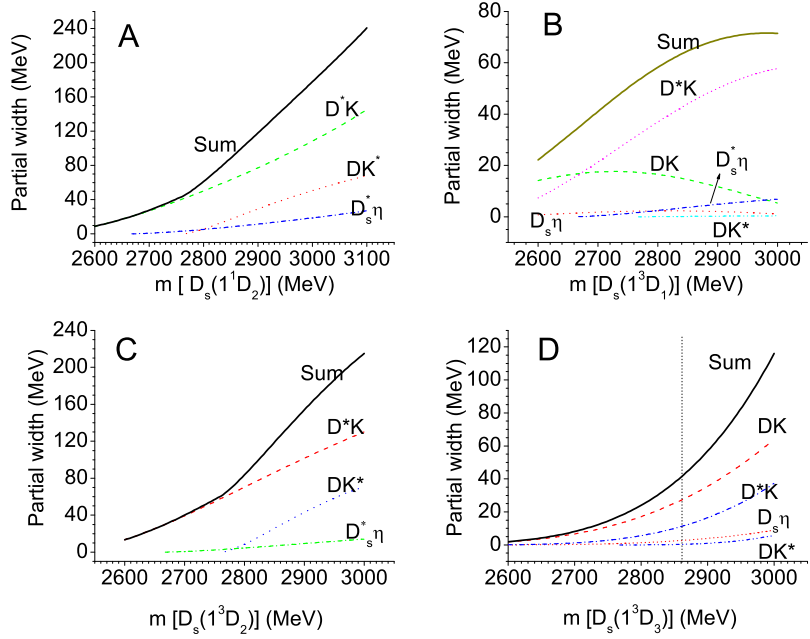} \caption{  The
partial widths of $D_s(1^1D_2)$, $D_s(1^3D_1)$, $D_s(1^3D_2)$ and
$D_s(1^3D_3)$ as functions of the mass. When we calculate the
partial width of the $DK^*$ channel, the $D$ meson is looked as
the emitted pseudoscalar meson in the $SU(4)$ case
\cite{Zhong:2007gp}. }\label{dsw}
\end{figure}
\end{center}
\end{widetext}

\subsection{Strong decays of $1D$ states}

Theoretical predicted masses of the $D$-wave excited $D$ and $D_s$
mesons are in a range of $2.8\sim 2.9$ GeV
\cite{Godfrey:1985xj,Matsuki:2007zz}. To see the decay properties
of those $D$-wave states, we plot their main decay partial widths
in their possible mass region in Figs. \ref{dw} and \ref{dsw} for
$D$ and $D_s$, respectively.

\subsubsection{ Excited $D$ mesons}

In Fig. \ref{dw}, decays of four $D$ wave states are presented. It
shows that the widths of $D(1^1D_2)$ and $D(1^3D_2)$ states are
dominated by $D^*\pi$ decay while the $D\pi/D\eta$ channels are
forbidden. At $\sim 2.8$ GeV, they have very broad widths larger
than 300 MeV. As $D^*$ dominantly decays into $D\pi$, such broad
widths imply that their dominant final states are $D\pi\pi$, and
it might be difficult to identify them in experiment. This may
explain why these states are still evading experimental
observations.

For $D(1^3D_1)$, $D^*\pi$ is also the main decay channel, but with
a much smaller width. In contrast, $D\pi$ dominates the decay of
$D(1^3D_3)$. With theory-suggested masses $m(D(1^3D_1))=2.82$ and
$m(D(1^3D_3))=2.83$ \cite{Godfrey:1985xj}, the total pionic decay
widths for $D(1^3D_1)$ and $D(1^3D_3)$ are predicted to be
\begin{eqnarray}
\Gamma(D(1^3D_1))\simeq 93 \ \mathrm{MeV},\\
\Gamma(D(1^3D_3))\simeq 130 \ \mathrm{MeV},
\end{eqnarray}
and the predicted ratios between the $D^*\pi$ and $D\pi$ widths
are
\begin{eqnarray}
R(D(1^3D_1))\equiv \frac{\Gamma(D\pi)}{\Gamma(D^*\pi)}\simeq 0.12,\\
R(D(1^3D_3))\equiv \frac{\Gamma(D\pi)}{\Gamma(D^*\pi)}\simeq 1.7.
\end{eqnarray}
For $D(1^3D_3)$, the dominance of $D\pi$ decay suggests that it is
relatively more accessible in experiment.

\subsubsection{ Excited $D_s$ mesons}

For $D_s(1^1D_2)$ and $D_s(1^3D_2)$, three important partial
widths of $D^*K$, $DK^*$ and $D_s^*\eta$ are presented in
Fig.~\ref{dsw}, and both states are dominated by the $D^*K$ decay.
It is interesting to see that the mass of $\sim$ 2.8 GeV is about
the threshold for $DK^*$. Although this decay mode is negligible
near threshold, it can become important in case that the masses
for these two $D$-wave states are above 2.8 GeV.

At 2.8 GeV, the total widths are dominated by the $D^*K$ mode,
which are
\begin{eqnarray}
\Gamma(D_s(1^1D_2))\simeq 61 \ \mathrm{MeV},\\
\Gamma(D_s(1^3D_2))\simeq 84 \ \mathrm{MeV}.
\end{eqnarray}
These results can guide a search for these two states around 2.8
GeV.

As shown in Fig.~\ref{dsw}, $D^*K$ and $DK$ are two dominant decay
modes for $D_s(1^3D_1)$ if its mass is around $2.8$ GeV, and the
predicted width is relatively narrow. Implications from such an
assignment will be discussed in subsection~\ref{2s-1d-mix}.

%[****** MORE DISCUSSION ABOUT THE MIXING RESULTS HERE]

Compared with the $D_s(1^3D_1)$ decay, the dominant decay mode of
$D_s(1^3D_3)$ state is $DK$ around 2.8 GeV. With a higher mass the
$D^*K$ channel becomes increasingly important. This feature fits
well the experimental observation for $D^*_{sJ}(2860)$, and makes
it a possible assignment for this state.

To be more specific, $D^*_{sJ}(2860)$ as a $D_s(1^3D_3)$ state,
its predicted width is
\begin{eqnarray}
\Gamma(D_s(1^3D_3))\simeq 41 \ \mathrm{MeV},
\end{eqnarray}
and the dominant decay mode is $DK$.  In comparison, the decays
via $DK^*$ and $D_s\eta$ are much less important (see the Fig.
\ref{dsw} and Tab.~\ref{width}). The ratio of $DK$ and $D^*K$ is
found to be
\begin{eqnarray}
R(D_s(1^3D_3))\equiv\frac{\Gamma(DK)}{\Gamma(D^{*}K)}\simeq 2.3 \
,
\end{eqnarray}
which is also consistent with the experiment~\cite{Aubert:2006mh}.
This assignment agrees with results of
Refs.~\cite{Colangelo:2006rq,Zhang:2006yj,Koponen:2007nr}.

Some models also suggested that $D^*_{sJ}(2860)$ could be a
$2^3P_0$ state \cite{vanBeveren:2006st,Zhang:2006yj,Close:2006gr},
for which only decay mode $DK$ and $D_s \eta$ are allowed. In our
model, a $2^3P_0$ state leads to decay amplitude
\begin{eqnarray}
&&\mathcal{M}(2^3P_0\rightarrow |1^1S_0\rangle
\mathbb{P})\nonumber\\
&=& i\frac{1}{4}\sqrt{\frac{1}{15}}g_I F
\frac{q^\prime}{\alpha}\left[ g^z_{10}\mathcal{G}q\frac{q^{\prime
2}}{\alpha^2} -\frac{1}{3}
g^z_{10}h q^\prime(1+\frac{q^{\prime 2}}{2\alpha^2})\right.\nonumber\\
&&\left.- 2\sqrt{2} g^+_{10} h q^\prime (7-\frac{q^{\prime
2}}{\alpha^2})\right]
\end{eqnarray}
with which its partial decay width to $DK$ is about $\Gamma=184$
MeV, and much broader than the experimental observation.

\subsection{ The $2^3S_1$-$1^3D_1$ mixing}\label{2s-1d-mix}

\begin{figure}[ht]
\centering \epsfxsize=8 cm \epsfbox{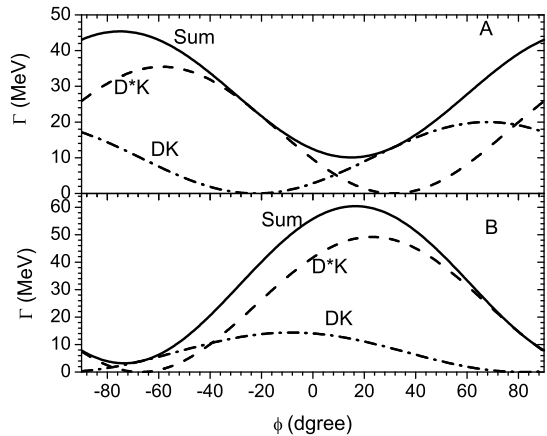} \caption{  The
partial widths of low vector (A) and high vector (B) as functions of
the mixing angle $\phi$. }\label{sdmix}
\end{figure}

In Ref.~\cite{Close:2006gr}, a mixing scheme between $2 ^3S_1$ and
$1^3D_1$ was proposed as a solution for the relatively broad
$D^*_{sJ}(2690)$, i.e.
\begin{eqnarray}\label{mix-s-d}
|D^*_{s1}(2690)\rangle & =&
\sin(\phi)|1^3D_1\rangle +\cos(\phi)|2^3S_1\rangle \ ,\nonumber\\
|D^*_{s1}(2810)\rangle &= & \cos(\phi)|1^3D_1\rangle -
\sin(\phi)|2^3S_1\rangle  \ ,
\end{eqnarray}
where an orthogonal state $D^*_{s1}(2810)$ was also predicted. The
mixing angle was found to favor $\phi=-0.5$ radians, i.e.
$\phi\simeq -27^{\circ}$. According to such a mixing scheme,
$D^*_{s1}(2810)$ will also be a broad state and dominated by $1
^3D_1$ configuration.

Taking the mixing scheme of Eq.~(\ref{mix-s-d}), we plot the
widths of $D^*_{sJ}(2690)$ and $D^*_{sJ}(2810)$ in terms of mixing
angle $\phi$ in Fig. \ref{sdmix}. From the figure, it shows that
if we take the mixing angle $\phi\simeq -27^\circ$ as predicted in
Ref.~\cite{Close:2006gr}, the decay modes of $D^*_{sJ}(2690)$ is
dominated by $D^*K$, which disagrees with the experimental
observation. Nevertheless, the predicted decay width of
$D^*_{sJ}(2690)$, $\Gamma\sim 25$ MeV, is underestimated by at
least a factor of $2$ compared with the data. If the $2S$-$1D$
mixing is small, e.g. $D^*_{sJ}(2690)$ is a pure $1^3D_1$ state,
the predicted decay width is $\Gamma\sim 42$ MeV, which is close
to the lower limit of the data. However, the ratio
$R=\Gamma(DK)/\Gamma(D^*K)\sim 0.8$ disagrees with the observation
that the $DK$ channel dominates the decay modes.

If we set $\phi\simeq 30^\circ$, which implies that the sign of
the spin-orbit splitting term is now negative to keep the correct
mass ordering, the $D^*K$ channel in $D^*_{sJ}(2690)$ decay will
be completely suppressed. This is consistent with the observations
except that the decay width $\Gamma\sim 15$ MeV is too small to
compare with the data.

For the high vector $D^*_{sJ}(2810)$, with $\phi\simeq \pm
30^\circ$, its width reads $\sim 40-60$ MeV and is dominated by
the $D^*K$. Meanwhile, its branching ratio to $DK$ is still
sizeable.

Our test of the property of $D^*_{sJ}(2690)$ in the $2S-1D$ mixing
scenario does not fit in the data very well.  To clarify the
nature of $D^*_{sJ}(2690)$, more accurate measurements of its
width and ratio $R=DK/D^*K$, and experimental search for the
accompanying $D^*_{sJ}(2810)$ in the $DK$ and $D^*K$ channels are
needed.

\section{Summary}\label{sum}

In the chiral quark model framework, we systematically study the
strong decays of heavy-light mesons in $\mathbb{M}\rightarrow
|1^1S_0\rangle \mathbb{P}$, $\mathbb{M}\rightarrow |1^3S_1\rangle
\mathbb{P}$, $\mathbb{M}\rightarrow |1^3P_0\rangle \mathbb{P}$,
$\mathbb{M}\rightarrow |1^1P_1\rangle \mathbb{P}$, and
$\mathbb{M}\rightarrow |1^3P_1\rangle \mathbb{P}$. By adopting
commonly used values for the constituent quark masses and
pseudoscalar meson decay constants, we make a full analysis of the
strong decays of all the excited $D^*, \ D_s^*, \ B^*$ and
$B_s^*$, and find that most available data can be consistently
explained in this framework. We summarize our major results as
follows.

\subsection{Excited $D$ mesons}

The calculated partial decay widths for the $D^*$ ($1^3S_1$),
$D^*_0(2400)$ as a $1^3P_0$ state and $D^*_2(2460)$ as a $1^3P_2$
state are in good agreement with the data in our model and support
their assignments as the low-lying excited $D^*$.

State mixing between the $1^1P_1$ and $1^3P_1$ is favored. With
the mixing angle, $\phi\simeq-(55\pm 5)^\circ$, which is
consistent with the prediction of Ref. \cite{Godfrey:1986wj}, the
narrow $D_1(2420)$ and broad $D'_1(2430)$ can be well explained as
mixing states between $1^1P_1$ and $1^3P_1$. Precise measurement
of the mass of the broad $D'_1(2430)$ is needed.

Our result shows that assigning $D^*(2640)$ to be a radially
excited $2^3S_1$ state can naturally explain its observation in
$D^{*+}\pi^+\pi^-$ final state, although the predicted width $\sim
34$ MeV still possesses some discrepancies with the data. A search
for the $D^*(2640)$ in the $D_1(2420)\pi$ channel is strongly
recommended.

Although the $2S$ and $1D$ excited $D$ states are still not well
established, we analyzed their partial decay widths in their
possible mass regions, which should be useful for future
experimental studies. The decay widths of $2S$ states turn to be
narrow, namely, at the order of a few tens of MeV. Our results
shows that $D(2^1S_0)$ is dominated by the $D^*\pi$ decay channel,
while both $D^*\pi$ and $D_1(2420)\pi$ are important for
$D(2^3S_1)$. Both $D(1^1D_2)$ and $D(1^3D_2)$ have very broad
widths $\gtrsim 300$ MeV, which may be difficult to identify in
experiment. The decay widths of $D(1^3D_1)$ and $D(1^3D_3)$ are
$\gtrsim 100$ MeV. The former has dominant decays into $D^*\pi$
and sizeable widths to $D\pi$ as well. For $D(1^3D_3)$, both
$D\pi$ and $D^*\pi$ are important.

\subsection{Excited $D_s$ mesons}

$D_s(2460)$ and $D_{s1}(2536)$ are identified as partners due to
mixing between $^1P_1$ and $^3P_1$. With mixing angle
$\phi\simeq-(55\pm 5)^\circ$, our predictions for $D_{s1}(2536)$
are in good agreement with the data. Note that $D_s(2460)$ is
below the strong decay threshold.

$D^*_{s2}(2573)$ is consistent with a $1^3P_2$ state. Its partial
decay width to $DK$ is dominant while to $D^*K$ is very small
($\sim 1$ MeV). This explains its absence in $D^*K$
channel~\cite{PDG}.

For those not-well-established states, by analyzing their decay
properties, we find that $D^*_{sJ}(2860)$ strongly favors a
$D_s(1^3D_3)$ state, while the $D^*_{sJ}(2632)$ and
$D^*_{sJ}(2690)$ cannot fit in any pure $D_s^*$ configurations.

For the unobserved $2S$ states, $D_s(2^1S_0)$  and $D_s(2^3S_1)$,
their decay widths are predicted to be an order of $\sim 15$ MeV,
and dominated by the $D^*K$ decay mode.

For the unobserved $1D$ states, $D_s(1^1D_2)$, $D_s(1^3D_2)$ and
$D_s(1^3D_1)$, their decay widths are of the order of $60\sim 80$
MeV at a mass of $\sim 2.8$ GeV. The $D_s(1^1D_2)$ and
$D_s(1^3D_2)$ decay are dominated by the $D^*K$ mode, while
$D_s(1^3D_1)$ is dominated by the $D^*K$ and $DK$ together.

\subsection{Excited $B$  mesons}

We also study the decay properties of the newly observed bottom
states $B_1(5725)$, $B^*_2(5829)$. Our calculations strongly
suggest that $B_1(5725)$ is a mixed $|P_1\rangle$ state, and
$B^*_2(5829)$ satisfies an assignment of $1^3P_2$.

The $B_J^*(5732)$, which was first reported by L3 collaboration,
can be naturally explained as the broad partner ($|P'_1\rangle$)
of $B_1(5725)$ in the $^1P_1$ and $^3P_1$ mixing scheme. Its
predicted width is $\Gamma(B_1')=219$ MeV, which is larger than
the PDG suggested value $\Gamma_{exp}=128\pm 18$ MeV. In contrast,
as a pure $1^3P_1$ state, its decay width is $\Gamma(B_1')=153$
MeV. Whether $B_J^*(5732)$ is a mixed state $|P'_1\rangle$, a pure
$1^3P_1$ state, or other configurations, should be further
studied.

The theoretical prediction of the mass for $B^*_0(^3P_0)$ meson is
about 5730 MeV.  Its decay into $B\pi$ has a broad width
$\Gamma(B^*_0)=272$ MeV according to our prediction.

\subsection{Excited $B_s$ mesons}

The two new narrow bottom-strange mesons $B_{s1}(5830)$ and
$B^*_{s2}(5839)$ observed by CDF are likely the mixed state $P_1$
and the $1^3P_2$ state, respectively, though their decay widths
and ratios are not given. $B_{s1}(5830)$ as a $|P_1\rangle$ state,
its predicted width and decay ratio are $\Gamma(B_{s1})\simeq
(0.4\sim 1)$ MeV and
$\Gamma(B_{s1})/(\Gamma(B_{s1})+\Gamma(B^*_{2s}))=0.02\sim 0.6$.
$B_s(5839)$ as the $^3P_2$ state, its decay width and width ratio
predicted by us are $\Gamma(B^*_{s2})\simeq  2 \ \mathrm{MeV}$ and
$\Gamma(B^*K)/\Gamma(BK) \simeq 6\%$.

The theoretical predictions for the $B^*_{s0}$ masses is about
5800 MeV. In our model it has a broad width $\Gamma(B^*_0)=227$
MeV. For $B'_{s1}$ if its mass is above the threshold of $B^*K$,
it will be a broad state with $\Gamma(B'_{s1})\simeq 149$ MeV.
Otherwise, it should be a narrow state.

It should be mentioned that uncertainties with quark model
parameters can give rise to uncertainties with the theoretical
results. A qualitative examination shows that such uncertainties can
be as large as $10-20\%$, which are a typical order for quark model
approaches. Interestingly, systematics arising from such a simple
prescription are useful for us to gain insights into the effective
degrees of freedom inside those heavy-light mesons and the
underlying dynamics for their strong decays.

%%%%%%%%%%%%%%%%%%%%%%%%%%%%%%%%%%%%%%%%%%%%%%%%%%%%%%%%%%%%%%%%%%%%%

\section*{  Acknowledgements }

The authors wish to thank F.E. Close and S.-L. Zhu for useful
discussions. This work is supported, in part, by the National
Natural Science Foundation of China (Grants 10675131 and
10775145), Chinese Academy of Sciences (KJCX3-SYW-N2), the U.K.
EPSRC (Grant No. GR/S99433/01), the Post-Doctoral Programme
Foundation of China, and K. C. Wong Education Foundation, Hong
Kong.

%\appendix
%%%%%%%%%%%%%%%%%%%%%%%%%%%%%%%%%%%%%%%%%%%%%%%%%%%%%%%%%%%%%%%%%%555

\end{document}